\documentclass[11pt]{prop} 
\usepackage[super,comma]{natbib} 
\usepackage{apjfonts} 
\usepackage{graphicx} 
\usepackage{amssymb,amsmath} 
\usepackage{color} 
\usepackage{enumitem} 
\usepackage{wrapfig} 
\usepackage{tikz} 

\usepackage{multirow} 
\usepackage{tabularx} 
\usepackage{array} 

\usepackage{hyperref} 
\hypersetup{
    colorlinks,
    linkcolor={blue!80!black},
    citecolor={blue!80!black},
    urlcolor={blue!80!black}
}

\usepackage{sidecap}
\sidecaptionvpos{figure}{c} 

\newcolumntype{L}[1]{>{\raggedright\let\newline\\\arraybackslash\hspace{0pt}}p{#1}}
\newcolumntype{C}[1]{>{\centering\let\newline\\\arraybackslash\hspace{0pt}}p{#1}}
\newcolumntype{R}[1]{>{\raggedleft\let\newline\\\arraybackslash\hspace{0pt}}p{#1}}

\definecolor{headcolor}{rgb}{0.65,0.65,0.65}
\usepackage{fancyhdr}
\renewcommand{\textfloatsep}{4mm}

\newcommand{\chandra}{\textit{Chandra}}

\newcommand{\xmm}{\textit{XMM-Newton}}

\newcommand{\bsf}{\sffamily\bfseries}

\definecolor{callout}{rgb}{0.25,0.40,0.85}
\definecolor{synergies}{rgb}{0.20,0.45,0.99}
\definecolor{methods}{rgb}{0.20,0.70,0.45}

\definecolor{calllem}{rgb}{0.20,0.45,0.99}
\definecolor{tabledef}{rgb}{0.95,0.95,0.95}
\definecolor{tablealt}{rgb}{0.77,0.80,1.0}
\definecolor{tablelem}{rgb}{0.80,0.85,1.0}
\definecolor{whitelem}{rgb}{1.0,1.0,1.0}
\definecolor{greenlem}{rgb}{0.7,1.0,0.7}

\usepackage{comment}
\usepackage[export]{adjustbox}

\usepackage{authblk}

\usepackage{color}

\newcommand{\etac}{$\eta$~Car}

\newcommand{\ms}{$M_{\odot}$}

\author[1,2]{Jeremy J. Drake}
\author[3]{David Cohen}
\author[4,5]{Michael Corcoran}
\author[4]{Maurice Leutenegger}
\author[2]{Kristina Monsch}
\author[6]{Ya\"el Naz\'e}
\author[7]{Lidia Oskinova}
\author[8]{Vallia Antoniou}

\affil[1]{Lockheed Martin, Palo Alto, CA 94304}
\affil[2]{Smithsonian Astrophysical Observatory, Cambridge MA 02138}{}
\affil[3]{Swarthmore College, Swarthmore, PA 19081}
\affil[4]{NASA/GSFC, Greenbelt, MD 20771}
\affil[5]{The Catholic University of America, Washington, DC 20064}
\affil[6]{FNRS/Univ. Li\`ege, All\'ee du 6 Ao\^ut 19c, B5C, 4000-Li\`ege, Belgium}
\affil[7]{Institute for Physics and Astronomy, University Potsdam, 14476 Potsdam, Germany}
\affil[8]{Texas Tech University, Lubbock, TX 79409}

\begin{document}

\baselineskip=13.2pt
\sloppy
\pagenumbering{roman}
\thispagestyle{empty}



\title{\textcolor{black}{\sf\huge \textcolor{blue}{\sl LEM GENERAL OBSERVER KEY SCIENCE}\\ \vspace{5mm} Addressing Outstanding Problems in the Physics of Massive Stars with the Line Emission Mapper X-ray Probe}\footnote{Corresponding author: Jeremy Drake (jeremy.1.drake@lmco.com)}}
\maketitle

\begin{tikzpicture}[remember picture,overlay]
\node[anchor=north west,yshift=2pt,xshift=2pt]%
    at (current page.north west)
    {\includegraphics[height=20mm]{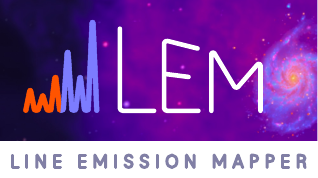}};
\end{tikzpicture}


\vspace*{-27mm}
\begin{center}
\begin{minipage}{17.5cm}
\hspace{-5mm}
\centering


\end{minipage}

\vfill
{\small
\phantom{${^52}$}~\textcolor{blue}{\bsf \href{https://lem-observatory.org}{lem-observatory.org}}\\
\phantom{${^52}$}~\textcolor{blue}{\bsf X / twitter: \href{https://www.twitter.com/LEMXray}{LEMXray}}\\
\phantom{${^52}$}~\textcolor{blue}{\bsf facebook: \href{https://www.facebook.com/LEMXrayProbe}{ LEMXrayProbe}}}
\vspace*{5mm}

\end{center}

\clearpage
\twocolumn


\section{SUMMARY}

The \textcolor{blue}{Line Emission Mapper} ({\it LEM}) is an X-ray Probe concept that, if appoved, will be slated for launch in the mid-2030s. The revolutionary capability of {\it LEM} lies in the its soft X-ray microcalorimeter array that achieves a resolution of 1~eV in the central array segment. In combination with a primary mirror comprising many pairs of thin monocrystalline silicon shells coated with either Ir or Pt, the instrument achieves an effective area exceeding 2000~cm$^2$ at 1~keV and averages $1600$~cm$^2$ over a 0.2-2~keV bandpass. This represents a gain of up to $\times$100 in effective area compared with the {\it Chandra} MEG, at similar spectral resolution. 

The gains that {\it LEM} provides for the study of massive stars are twofold: the ability to obtain high-resolution X-ray spectra for much fainter objects; and the sensitivity to perform time-dependent studies of brighter objects. The former quality enables study of objects at order of magnitude greater distances than have been accessible to date, opening up an immense volume of space within which suitable targets can be found. Since 70\%\ of the observing programs of the Probe Class missions will be dedicated to General Observer access, there will be vast scope for making transformative progress in our understanding of the energetic processes of early-type stars and the resulting feedback into their environments.

In this paper, we outline the main points of the scientific case for high-resolution X-ray spectroscopy of early-type stars with {\it LEM}. There are several major outstanding problems that {\it Chandra} and {\it XMM-Newton} have opened up, or were unable to address. These have much wider importance than the stellar physics itself: early-type stars are important drivers of chemical enrichment, energy recycling and feedback in galaxies, through both their winds and their end of life supernova explosions.

X-rays from single early-type stars are generated by shocked plasma resulting from instabilities in their line-driven winds, but the winds suffer from  inhomogeneities and clumping and only high-resolution X-ray spectra can resolve the spectral line profiles that enable accurate mass loss rate and feedback energy assessment. 
These line-driven winds are expected to be highly dependent on the metal abundances responsible for the UV absorption lines, and {\it LEM} is capable of observing the brighter O stars in the  metal-poor environments of the Magellanic Clouds. 

A small fraction (7\%) of OB stars are strongly magnetic and their winds are channeled by the field, leading to interacting regions and further shock heating that can be diagnosed at high spectral resolution. Since stellar multiplicity is a strongly increasing function with stellar mass, a large fraction of early-type stars are in multiple systems where their supersonic winds collide and interact. The spectral line profiles and changes through the orbit provide diagnostics of the wind interaction regions, and insights into wind acceleration and radial velocity profiles and mass loss rates. 

These and other important problems at the heart of understanding X-rays from high-mass stars will become tractable with {\it LEM}.

\section{INTRODUCTION}
\label{s:intro}

X-ray emitting O and early B stars have masses of approximately $10M_\odot$ or more. 
They undergo rapid evolution, eventually transforming into Wolf-Rayet stars, with typical lifespans of only 4-10 Myr. They play a crucial role in the energetics, chemical enrichment and evolution of their host galaxies. Their intense ultraviolet emission makes them the primary source of ionization of their local interstellar media. Throughout their lifetimes, they inject vast amounts of radiative and mechanical energy into their surroundings that can both trigger and suppress star formation. Ultimately, these stars meet their end in core-collapse supernova explosions (SNe), enriching their surroundings with metals, and energizing their environments with kinetic energy. 

Massive OB-type stars have powerful UV radiation fields that drive rapidly expanding winds at speeds exceeding 1000~km~s$^{-1}$ via spectral line scattering. These winds were initially seen in sounding rocket observations \citep{1967ApJ...147.1017M} and were one of the first major discoveries of space-based astronomical instrumentation.  The wind properties were explored with observations made by the {\it Copernicus} satellite in the mid-1970s\cite{1976ApJS...32..429S}, which revealed far-UV spectra characterised by P~Cygni line profiles. 

Mass loss from these radiatively-driven stellar winds strongly influences the evolution of the  star to a SNe and a black hole or neutron star, and  
stellar winds deposit both mass and energy into the local environment, driving enrichment of the local ISM as this outflow evolves in strength and chemical composition. 
The evolutionary pathways of massive stars are strongly affected by the integrated
mass loss over their lifetime. Understanding high-mass stars and their mass loss is not only important in its own right, but is also key to understanding the dominant source of stellar feedback in galaxies and its different aspects, including star formation, wind bubbles, supernovae, and galactic evolution, as well as the origin of compact objects. The X-ray emission from high mass stars is central to solving these outstanding problems.

\begin{figure}[t]
\begin{center}
\includegraphics[width=0.49\textwidth]{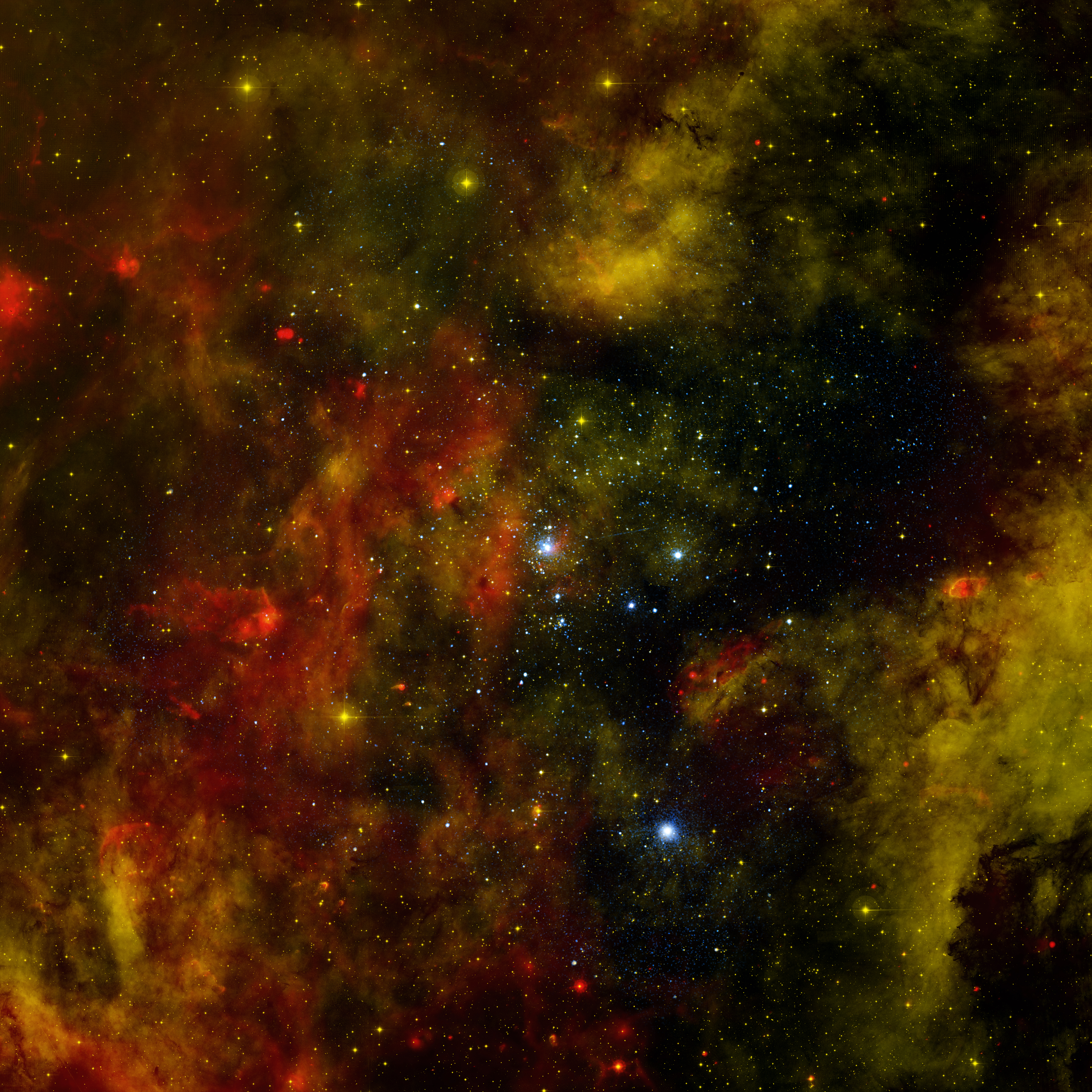} 
\end{center}
\caption{The central region of the nearby ($\sim 1.4$~kpc) massive Cygnus OB2 Association containing a total stellar mass of  approximately 30,000$M_\odot$\cite{Wright2010}, compiled from observations by {\it Chandra} in X-rays (blue), Isaac Newton Telescope in visible light (yellow), and {\it Spitzer} in the infrared (red). The region is estimated to range in age from 3-5~Myr or so, with no known supernova having occurred to date. As such, the stellar feedback into the region is presently driven by the winds from the OB stars that also dominate the bright X-ray source population. It was the brightest of these that were the first massive stars detected in X-rays.\cite{Harnden1979} The image covers approximately 1.5$\times$1.5 degrees on the sky. Based on Refs.~\citenum{Wright2014,Guarcello2015}.
 }
\label{f:cygob2}
\end{figure}

X-rays from high-mass stars were first discovered serendipitously by the {\it Einstein} observatory from two different observations made consecutively on 1978 December 14-17. One was targeting the Cygnus X-3 high-mass X-ray binary and detected X-ray bright O stars belonging to the relatively nearby ($\sim 1.4$~kpc) massive Cygnus~OB2 Association star forming region \cite{Harnden1979}. The other was probing the origin of X-rays detected by earlier missions in the Carina star-forming region, and was able to associate X-rays with several O-type stars and with the notorious luminous blue variable $\eta$~Carina\cite{1979ApJ...234L..55S} (see also Sect~\ref{sec:cwb}). Figure~\ref{f:cygob2} illustrates a multi-wavelength image of Cygnus~OB2 with X-rays obtained by {\it Chandra} shown in blue.\cite{Wright2014,Guarcello2015} The brightest X-ray sources are the OB stars that were first seen by {\it Einstein} and presently dominate the feedback into the region via their energetic winds. 

The finding of X-rays from high-mass stars was somewhat unexpected since (except in rare instances) massive stars do not possess strong, dynamo-driven magnetic fields which produce X-ray emission in the Sun and lower-mass stars.  Still, the presence of hot plasma in OB star winds had been inferred\citep{Cassinelli1979} based on analysis of the UV spectra of O-type supergiants.  Subsequent observations by {\it Einstein}, {\it ROSAT} and other X-ray telescopes found that the X-ray luminosity of O-type and early B-type stars is typically linked to their bolometric luminosity, following an approximate relationship of $L_X/L_{bol} \approx 10^{-7}$, with stars of type later than mid-B being generally X-ray dark  \citep{1997A&A...322..167B, 1980ApJ...239L..65L}. 

The wind origin of the X-rays from massive stars was confirmed soon after the launch of \chandra\/ and \xmm\/ from Doppler broadening of X-ray emission lines, showing that OB star X-rays are due to wind shocks \citep{2001ApJ...554L..55C,Kahn2001}. The line-deshadowing instability (LDI)\cite{1980ApJ...241..300L}, intrinsic to line-driven flows, produces wind structure, variability, and shocks, which heat a small fraction of the wind to X-ray emitting temperatures \citep{1988ApJ...335..914O}. It is commonly thought that photospheric variability, perhaps caused by subsurface convection zones, enhances the shocks and X-rays beyond what the LDI itself can produce 
\citep{Massa2019ApJ...873...81M,2022MNRAS.513.2663G}. However, many questions about both the X-ray production and the winds themselves remain open to exploration with a new generation of more sensitive X-ray spectrometers.

%

In addition to the open questions about X-ray production in effectively single OB stars via the LDI, there are unsolved issues related to colliding wind massive binaries\cite{1976SvAL....2..138C,2021ApJ...923..191P}, to the 7\%\ or so of OB stars with strong dipolar magnetic fields\citep{2016MNRAS.456....2W} that channel the wind flow, and to the case of Wolf-Rayet stars whose X-ray emission remains poorly understood\cite{Oskinova:2012lr}. 

Here, we review some outstanding issues in the physics of high-mass stars and outline  progress that could be made with the {\it Line Emission Mapper} ({\it LEM}) X-ray Probe. In terms of effective area {\it LEM}, has 
order of magnitude or greater performance advantage compared with the diffraction gratings on {\it Chandra} and {\it XMM-Newton}, extending the horizon for high-resolution X-ray spectroscopy to much fainter sources and opening the opportunity for time-dependent study of brighter objects. 
We first note some key outstanding problems in the physics of massive stars, then note the salient characteristics of {\it LEM} that could lead to new breakthroughs. We then address the ways in which {\it LEM} could lead to progress in solving outstanding problems.


\section{KEY OUTSTANDING PROBLEMS IN HIGH-ENERGY HOT STAR PHYSICS}
\label{s:outstanding}


As we describe below, unsolved problems in the high energy physics of massive stars are all related in one way or another to understanding the their winds, or the interaction of their winds with companion stars, and/or their surroundings.
High-resolution X-ray spectroscopy with next-generation instruments 
is needed to resolve the current list of outstanding issues, and to reveal new ones.

\paragraph{Clumping and Mass Loss Rates} 

The wind mass-loss rates of OB stars are ``high'' in the sense that they amount to a significant fraction of the stellar mass over the stellar lifetime, and consequently have evolutionary implications. However, they are difficult to measure accurately, not least because the shocks distributed throughout the wind that produce the X-rays also make the wind clumpy. Clumping affects density-squared diagnostics, making optical, IR, and radio mass-loss rate determinations uncertain \citep{2006A&A...454..625P} and also affects ionization balance which makes UV mass-loss rate diagnostics uncertain \citep{2006ApJ...637.1025F}. Accurate mass-loss rate measurements and characterization of clumping properties are two interconnected challenges that only high-resolution X-ray spectroscopy can address \citep{Owocki2001,Owocki2006,Oskinova2006,Cohen2010,Sundqvist2012,Leutenegger2013,Cohen2014,2013A&A...551A..83H,2022MNRAS.513.2663G}.

\paragraph{Metallicity} An important  issue is  the dependence of wind properties, including mass loss rate, on stellar metallicity. Since wind driving is due to metal lines, wind properties should depend sensitively on stellar metallicity. This dependence is especially important for our understanding of the chemical evolution of the ISM via stellar winds in the early Universe. Interestingly, 
recent studies of X-ray emission\citep{Crowther2022MNRAS.515.4130C, Oskinova2012,2014ApJS..213...23N} from massive stars in the Small Magellanic Cloud (SMC) and Large Magellanic Cloud (LMC) -- environments with average metallicities of 0.2 and 0.5 times solar, respectively -- appear equally efficient in generating X-ray emission and exhibit similar ratios of bolometric to X-ray luminosity, $L_X/L_{bol}$, as Galactic OB stars\cite{Oskinova2005,Naze2009,2011ApJS..194....7N,2015A&A...580A..59R,Nebot2018} (Sects.~\ref{s:singles},\ref{s:othergals}).

\paragraph{Magnetically-Confined Winds} Understanding the magnetospheric structure of the 7\% of OB stars that are strongly magnetic is a key issue that X-ray spectroscopy can address. Of the (rather) small sample studied thus far, the shocked plasma temperature is generally lower than models predict, some cases are too luminous, and the X-ray modulations appear puzzling \cite{2014ApJS..215...10N,2016AdSpR..58..680U,2016ApJ...831..138N} 
X-ray line spectrometry provides better understanding of the magnetospheric structure to improve our understanding of the origins of the fields and the connection to magnetars and heavy stellar mass black holes\cite{2017MNRAS.466.1052P} (Sect.~\ref{s:magnetic}). 

\paragraph{Colliding Winds} The stellar binary fraction is observed to increase with increasing stellar mass \citep{DucheneKraus13}, and consequently a large fraction of early-type stars are found in multiple systems. Colliding winds in such systems give rise to particularly bright and hard X-ray emission \citep{Stevens1992, Rauw2016}. The behavior of X-ray emission through orbital phase should depend on the orbital eccentricity and the wind radial density and velocity profiles. In turn, these parameters depend on details of the wind driving. 
Thus colliding wind binaries present a unique laboratory for the determination of stellar wind properties via  sensitive X-ray line spectrometry, as the X-ray line profiles are sensitive to the wind flow dynamics and density\cite{2016NewA...43...70R,2021A&A...646A..89M,2022MNRAS.513.6074M} (Sect.~\ref{sec:cwb}). 

\subsubsection{X-rays from Wolf-Rayet Stars} Both single Wolf-Rayet (WR) stars and WR stars in colliding wind binaries produce X-rays. While the colliding wind systems can largely be understood under that paradigm (although with notable exceptions\cite{2009A&A...508..805G}), the wind properties of WR stars are significantly different from those of less evolved OB stars, and X-ray emission mechanisms for single WR stars is far from understood\citep{2016AdSpR..58..739O}. 
X-ray emission line spectrometry is the best diagnostic of X-ray emission mechanisms, and needed to understand the location and dynamics of the hot, X-ray emitting gas (Sect.~\ref{s:wr}).

\paragraph{Be Star X-rays} The origin of the X-ray emission from a significant ($\sim$10\%\cite{2018A&A...619A.148N,2023MNRAS.525.4186N}) fraction of Be star binaries also poses a puzzle. These are the $\gamma$~Cas systems, that exhibit unusually hard and bright X-ray emission compared with other Be stars \citep{Smith2016}. The origin of this more intense X-ray emission is not known and is actively debated \citep{Postnov2017, Naze2022, Gies2023} (Sect.~\ref{s:bestars}).

\paragraph{Exoplanets around hot stars}
Recent detections of planets around hot stars such as b Centauri (AB)b \citep{Janson+2021}, Mu2 Scorpii b \citep{Squicciarini+2022} or V921 Sco b \citep{UbeiraGabellini+2019}, present a puzzle for planet formation theory. The strong UV, EUV and X-ray radiation fields of these stars, in combination with their rapid disk evolution, suggest that planets should either not have time to form, or else will be rapidly evaporated once formed. {\it LEM} will provide new insights into the very soft X-ray radiation fields of high mass stars which will help in understanding how their planets survive (Sect.~\ref{s:exoplanets}).

\section{LEVERAGE FROM HIGH-RESOLUTION X-RAY SPECTROSCOPY WITH {\it LEM}} 
\label{s:leverage}

{\it LEM} features a large microcalorimeter array/IFU, with a 30$\times$30' field of view, 10" angular resolution, and an effective area peaking at over 2000~cm$^2$, with an average through the band of 1600~cm$^2$.\cite{Kraft2022} This represents more than an order of magnitude increase compared with the {\it XMM-Newton} Reflection Grating Spectrometer, and a factor of 20-100 times the effective area of the {\it Chandra} Medium Energy Grating. 

The {\it LEM} microcalorimeter is tuned to lower energies than that of the X-ray Integral Field Unit (X-IFU)  instrument on the planned European Space Agency-lead {\it Athena} mission, and will achieve an energy resolution of 1~eV in the 0.2-2 keV band in the central 7' part of the array, with the outer parts of the array reaching energy resolutions of 2~eV. The central part of the array would be used for point sources, such as the early-type stars considered here, in a guest observer phase, and the energy resolution corresponds to a resolving power $E/\Delta E=1000$ at 1~keV (12.4~\AA).
X-ray spectra from wind shocks in early-type stars 
are sensitive to 
the distribution and dynamics of shocked and cooled post-shock wind plasma. Spectral line profiles provide sensitive constraints on the location of the hot gas in in the wind, 
and of the chemical composition of the shocked gas, largely independent of photoionization biases.

Many types of localized wind structures 
can be diagnosed at X-ray wavelengths like  wind confinement of hot gas in magnetic stars, corotating interaction regions, and the hot "bow-shock" structures of colliding wind systems. 

{\em Only a handful of X-ray bright massive stars have been observed at high-resolution in X-rays, and even fewer have been monitored for tell-tale variations in their spectra.}

{\it LEM} will also provide crucial information on the soft X-ray environment near OB stars. 
Understanding this environment is important to our understanding of possible planetary formation near OB stars, since the soft X-ray environment plays an important role in 
heating an dissipating the gaseous protoplanetary disk\citep{Ercolano+2009, Ercolano+2021} as well as exoplanet atmospheres \citep[e.g.][]{Murray-Clay+2009, OwenJackson2012}.

\subsection{THE DIAGNOSTIC POTENTIAL OF HE-LIKE IONS FOR HOT STARS}
\label{s:helikes}

\begin{figure}[t]
\begin{center}
\includegraphics[width=0.49\textwidth]{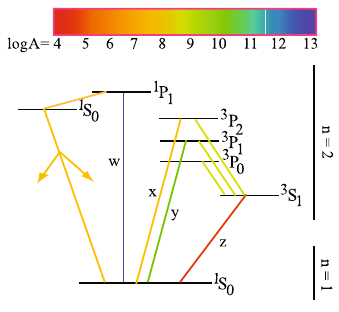} 
\includegraphics[width=0.49\textwidth]{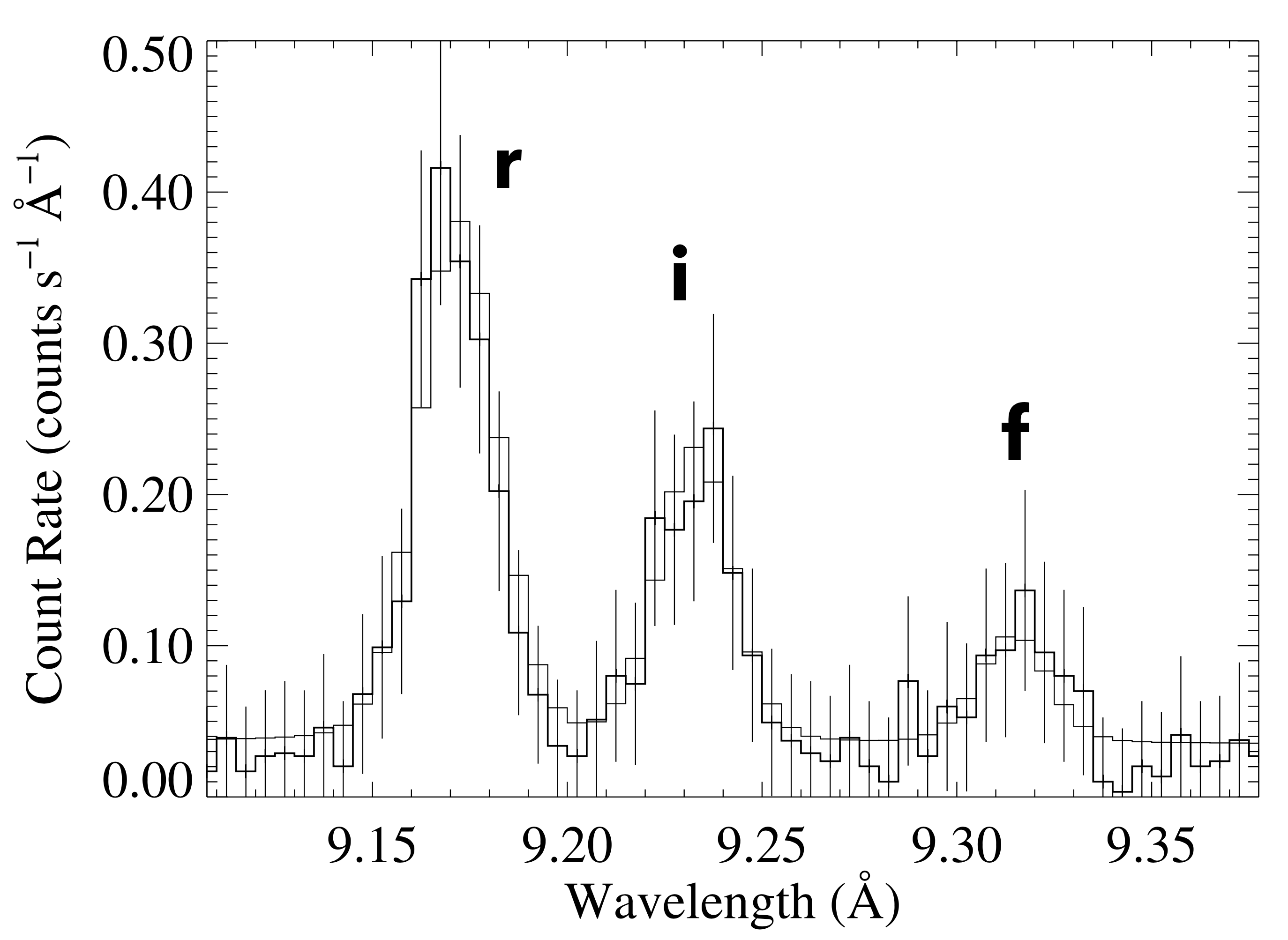} 
\end{center}
\caption{Top: A simplified diagram of energy levels for He-like ions from Ref.~\citenum{Foster.etal:10} illustrating the resonance ($w$ sometimes referred to as $r$), intercombination ($x$ and $y$, or sometimes $i$), and forbidden ($z$, or sometimes $f$) line transitions. The relative A-values are represented using different colors. Bottom: The Mg~XI He-like complex observed in the B0.2~V$ star \tau$~Sco using the {\it Chandra} HETG. In the low density and low radiation field limit, $f>i$ typically by a factor of 2 or more. Here, $i$ is significantly enhanced relative to $f$ by FUV pumping of $^3S_1$-$^3P_{1,2}$ due to the intense photospheric radiation field. From Ref.~\citenum{Cohen2003}.
}
\label{f:helikes}
\end{figure}

High-resolution spectroscopy can be utilized to gain insights into the spatial structure and location of source X-ray emitting regions. The emitting power of a spectral line excited by electron collisions depends on the product of the plasma density squared and the volume the plasma occupies. However, there are other density dependencies in spectral line strengths that arise from collisional excitation equilibria within ions and associated changes in relative level populations involving metastable and forbidden decay routes. The most well known of these in X-ray spectroscopy are the ratios of forbidden to intercombination lines in He-like ions that are sensitive to plasma density,\cite{Gabriel1969,Blumenthal1972} as illustrated in Figure~\ref{f:helikes}. These lines have been commonly used to estimate densities and therefore the emitting volumes of the source plasma, providing clues to its structure. 

Also relevant for early-type stars is the sensitivity of the 
The ratio of forbidden to intercombination line strengths to the local intensity of the UV radiation field\cite{Blumenthal1972,Cohen2003}.  The transitions $^3P_{1,2}$-$^3S_1$ lie in the far ultraviolet for the commonly observed He-like ions of N, O, Ne and Mg, and if radiative excitation from the $^3S_1$ level to the $^3P_{1,2}$ levels becomes comparable to the direct decay rate of $^3S_1$ to the ground state, further radiation intensity increase leads to an increase in the intensity of the intercombination line, $i$, while the forbidden line intensity, $f$, decreases. Consequently, measuring the $f/i$ ratio has the potential to provide information about the radiation field and the radius of the X-ray emitting plasma by considering its sensitivity to the mean intensity and radiation dilution factor.\cite{Kahn2001,Cohen2003,Raassen2008,Hole2012}  Variations in the predicted $f/i$ line ratio resulting from stellar pulsations and variations in wind properties have also been explored.\cite{Hole2012,Ignace2019}

\section{X-RAYS FROM SINGLE HOT STARS}
\label{s:singles}

The winds of O and B stars are driven by radiation pressure in spectral lines from intense photospheric UV emission \cite{1967ApJ...147.1017M,1976ApJS...32..429S}. They contribute significant mass, energy, and momentum to their environments, and are important sources of feedback early in an episode of star formation (see, e.g., Figure~\ref{f:cygob2}). They are also the progenitors of the core-collapse supernovae (CC SNe) that are the primary sources of stellar feedback, with their evolutionary pathways strongly affected by integrated mass loss over their lifetimes. The quantitative calibration of their wind properties is thus instrumental in constraining their direct contribution to feedback and their indirect effect on evolutionary pathways leading to feedback from CC SNe \citep{2008A&ARv..16..209P}.

O and very early B stars are strong sources of soft X-ray emission, arising from shock-heated wind plasma\cite{1988ApJ...335..914O, 1997A&A...322..878F}.  The approximately linear correlation between bolometric and X-ray luminosities means the earliest O star in young star clusters is often the source with the highest X-ray luminosity\cite{Oskinova2005,Naze2009,2011ApJS..194....7N,2015A&A...580A..59R,Nebot2018}. X-ray spectroscopy has proved to be a powerful tool for constraining stellar wind models and their predictions of both wind heating and overall wind structure. 

\begin{figure*}
	\includegraphics[angle=0,width=0.98\textwidth]{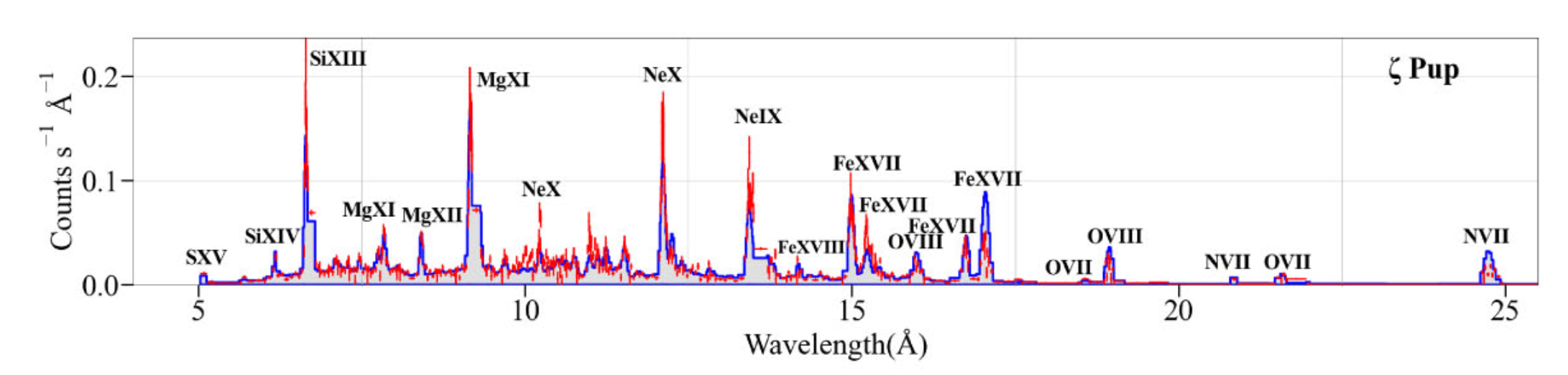}
	\includegraphics[angle=0,width=0.98\textwidth]{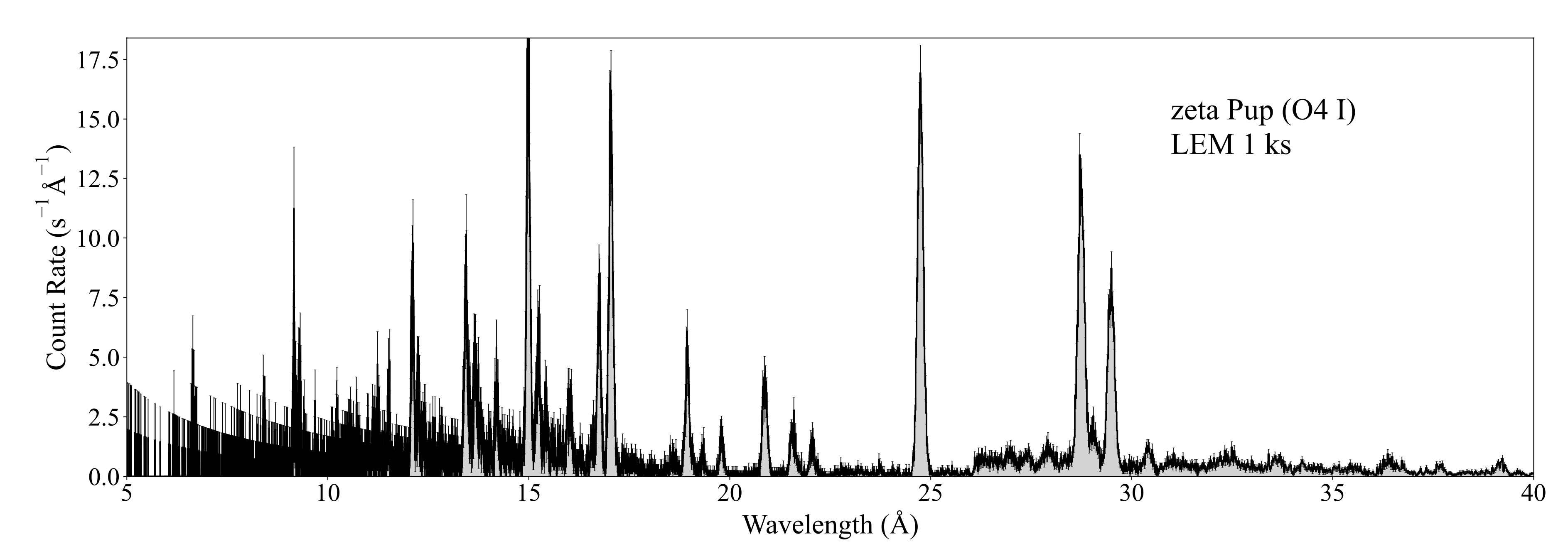}
	\includegraphics[angle=0,width=0.98\textwidth]{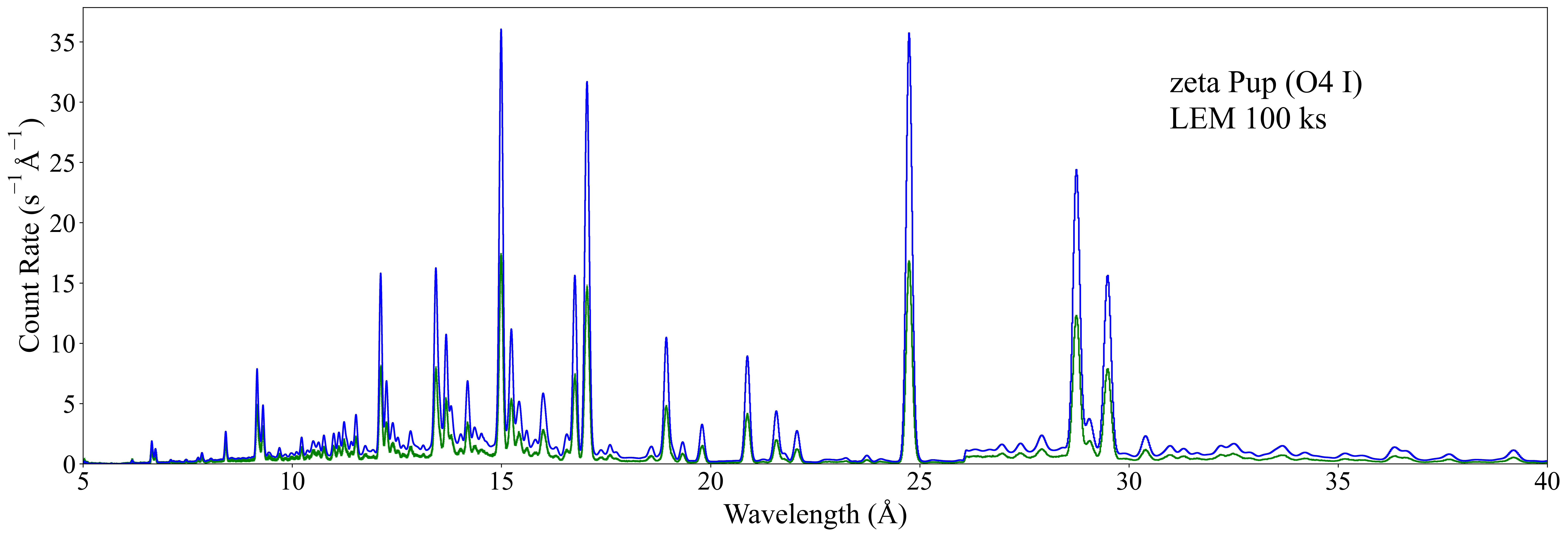}
	\caption{The {\it Chandra} MEG cycle 1 spectrum of $\zeta$~Pup obtained on 2000 March 28 is presented in a grouped representation  (top) along with the best-fit multi-temperature emission and wind absorption model (Cohen et al.\ 2021). The exposure time is 67 ks. Below that is a simulation of an {\it LEM} observation of only 1 ks, using the best-fit model from the top panel (and added Poisson noise). The simulated {\it LEM} spectrum is shown at its intrinsic binning and has signal-to-noise superior to the much longer Chandra MEG observation. In the bottom panel, we show a 100 ks simulation of the same star but with two different values of the wind mass-loss rate (differing by a factor of two -- the lower one in blue, the higher one in green). The error bars are too small to see. With this exquisite data quality, the mass-loss rate can be determined to better than 1 percent, formally.
		}   
	\label{fig:zetaPup_Chandra}
\end{figure*}

\begin{figure*}[ht]
	\includegraphics[angle=0,width=0.49\textwidth]{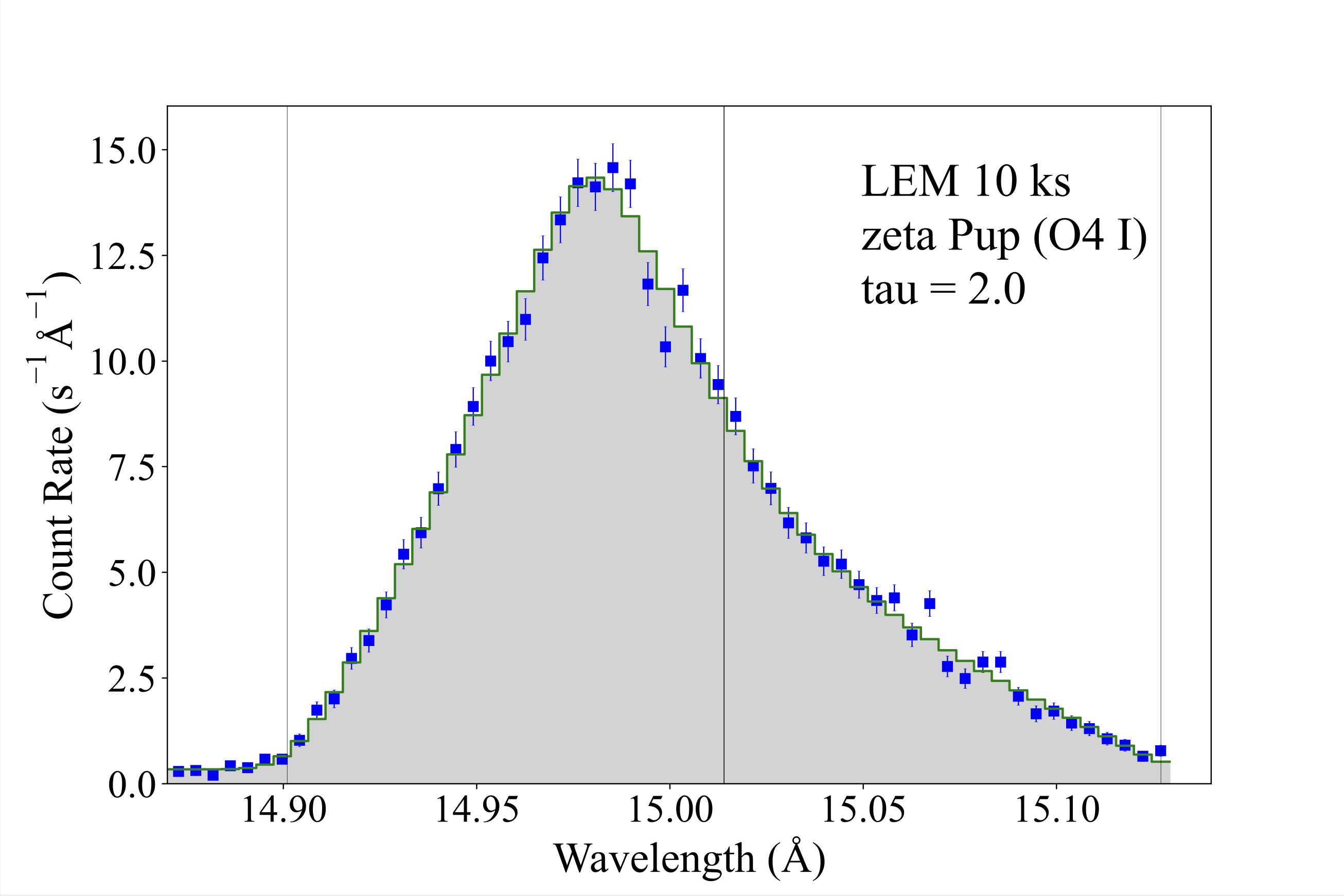}
	\includegraphics[angle=0,width=0.49\textwidth]{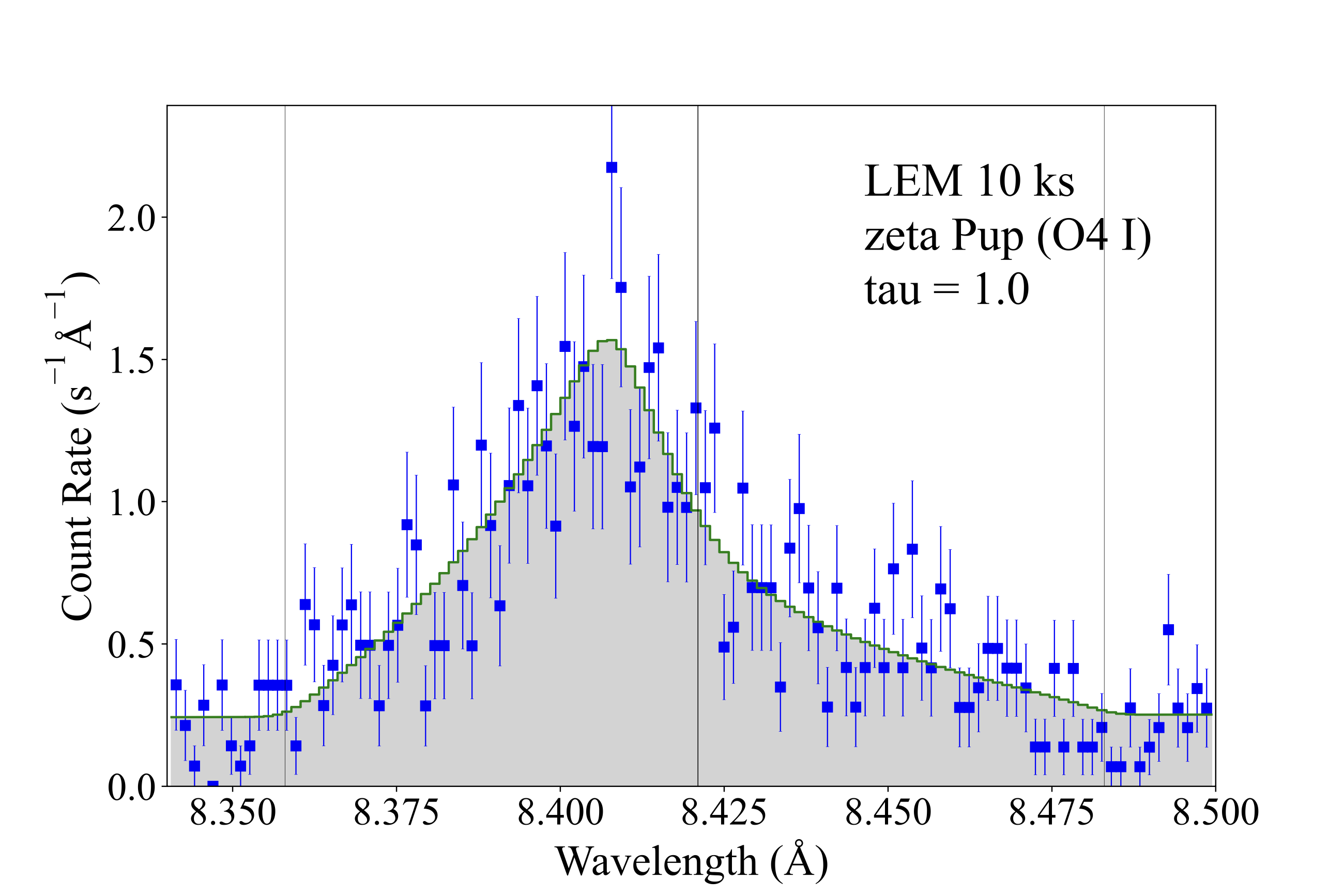}
	\caption{Simulations of two Doppler-broadened emission lines from the wind of zeta Puppis: Fe XVII at 15.01 \AA\ (left) and Mg XII at 8.421 \AA\ (right). The different optical depths affect the line shapes and can be used to measure the wind mass-loss rate. The higher signal-to-noise ratio of the iron line means its optical depth can be constrained to a few percent, while the optical depth of the shorter wavelength magnesium line can only be constrained to a few tens of percent. 
		}   
	\label{fig:zetaPup_lines}
\end{figure*}

No more than two dozen Galactic OB stars have been observed with the grating spectrometers on {\it Chandra} and {\it XMM-Newton} with sufficient exposure times to yield high-quality data, but a significant amount of key information has been derived from this limited sample of high-resolution X-ray  spectra, including\cite{Leutenegger2006,2013A&A...551A..83H,2014MNRAS.444.3729C, 2022MNRAS.513.2663G, 2022MNRAS.513.1609C}: (1) temperature distributions and radial distribution of hot plasma within the wind that constrains models of wind shocks and X-ray production; (2) wind and shock kinematic properties via Doppler-broadened and resolved emission line profiles; (3) wind mass-loss rates from emission line profile asymmetries and broadband absorption signatures; and (4) abundances, including often extremely altered N/O ratios. In Fig.\ \ref{fig:zetaPup_Chandra}, the Chandra grating spectrum of the O supergiant $\zeta$ Pup is shown, along with an exquisite simulated {\it LEM} spectrum representing just a 1 ks exposure, and in Fig.~\ref{fig:zetaPup_lines}, {\it Chandra} data and an {\it LEM} simulation of key line profile diagnostics are shown. 

Mass-loss rates and abundances are particularly important diagnostics for studies of feedback, and diagnostic techniques in other wavelength bands for both of these are limited by systematic modeling errors. Mass-loss rate diagnostics are compromised by wind inhomogeneities (clumping), which can affect diagnostics that scale with the square of density (radio emission, H$\alpha$ recombination), but also even those that depend on resonance lines (UV P Cygni profiles) through non-monotonic velocity fields. Abundances are determined by global spectral modeling of UV and optical emission and absorption lines, which are compromised by uncertainties in atomic data and systematic errors in complex radiative transfer codes. X-ray diagnostics are much simpler, are not compromised by these effects, and are arguably the most reliable tools at our disposal.  High-quality diagnostics and discriminants of competing models can be obtained by careful line profile analysis including the possibility of plasma emission over different locations depending on temperature\cite{2013A&A...551A..83H,2022MNRAS.513.2663G,Cohen2010,Cohen2014}.


Progress depends on expanding the number of stars studied using high-resolution, sensitive X-ray spectrometers, both in the Galaxy at solar metallicity, and in the Magellanic Clouds at subsolar metallicity. With its large effective area, {\it LEM} would enable {\it detailed study} of the high-resolution X-ray spectra from hundreds of OB stars. Some could be serendipitously observed by {\it LEM} since early-type stars, being drivers of feedback, occur frequently in star forming regions, which will be extensively studied by {\it LEM}. 

Line-driven winds are not limited to massive OB stars and their descendants. While not under the main topic of this paper, we note that low-mass subdwarfs of the OB spectral type also display an intense UV emission, and may eject significant outflows. Using the previous generation of facilities, they were detected to emit X-rays similar to those of massive OB stars - in particular obeying the canonical $L_{\rm x}/L_{\rm bol}$ relationship and corresponding to plasma at tepid temperatures\cite{2016AdSpR..58..809M}. Clearly, the same physical process seems to be at play, but none of those stars have been investigated at high resolution. This will be an "ultimate" test of the embedded wind shocks thought to power the X-rays, taking that concept to the most extreme conditions: this will thus be a sensitive test of models, allowing to build a robust physical understanding of the physics involved.

In a more general way, since X-rays of massive stars are born in the winds, the presence of wind features will necessarily affect the X-ray emission properties. This has been discussed above for embedded wind shocks, but there are additional possibilities. One such case is corotating features (CIRs)\citep{Mullan1984}. Rotational modulations have been detected for massive OB stars at various wavelengths. Photometric variations in the optical are thought to be linked to bright spots \cite{2018MNRAS.473.5532R}. Such spots would then locally drive a different outflow, which interacts with the regular wind to produce large-scale wind features called "corotating interaction regions". Their signatures are ubiquitous in UV spectroscopy \citep{Kaper1996, Massa2019ApJ...873...81M}. 

The previous generation of X-ray facilities have revealed the counterparts of these corotating interaction regions at high energies. Stars like 
$\zeta$~Oph \citep{Oskinova2001},  $\lambda$~Cep\cite{2015A&A...580A..59R}, or $\xi$~Per \citep{Massa2019ApJ...873...81M} indeed display X-ray variations compatible with the timescales of the optically- and UV-detected modulations. The best example is $\zeta$~Pup, where analysis of 1~Ms observations made by both {\it Chandra} and {\it XMM-Newton} directly detected the periodicity of the X-ray changes and demonstrated that it was the same as that detected in simultaneous optical photometry\citep{2018A&A...609A..81N, Nichols2021}. However, most of this was done with broad-band data. 
Unfortunately, the low sensitivity of these data did not allow meaningful constraints on the modulation of the high-resolution spectra. 

With {\it LEM}, the large effective area means that line profiles and how they vary can be studied in dozens of OB stars. 
X-ray line variability studies with {\it LEM} compared to optical and UV variations can constrain the relation between photospheric emission and the ambient unshocked and shocked wind.  
The X-ray line studies can help distinguish
changes in X-ray absorption along the line-of-sight from  changes in flow velocities or density structures.  
A particularly interesting study that will be enabled by {\it LEM} is to determine how efficiently stellar pulsations propagate from the  photosphere into the wind\citep{2017A&A...608A..54C,2014NatCo...5.4024O}.


\section{MAGNETIC OB STARS}
\label{s:magnetic}

In magnetic massive stars, if the field is strong enough, it can confine and trap the wind and significantly reduce the star's mass-loss rate, affecting core collapse progenitor masses and thus SN explosion characteristics, as well as potentially producing more massive black holes \citep{2017MNRAS.466.1052P}. Winds and supernovae are key massive-star drivers of feedback and better understanding the role of magnetic fields in massive stars will improve our understanding of these feedback inputs. 

Only 7\% of OB stars  have strong (few 100 to $10^4$~G), large-scale, and stable (likely fossil) magnetic fields with significant magnetospheres fed by their radiation-driven winds. The wind-feeding leads to shock heating and X-ray emission, which is strong because the magnetically-channeled wind is fast and the confined magnetospheric plasma is basically stationary. Because it is rare for the magnetic and rotation axes to be aligned, observables, including X-rays, are often modulated by stellar rotation\cite{2014ApJS..215...10N,2015MNRAS.452.2641N,2023MNRAS.521.2874R}. 
Time-variable X-ray spectra provide important information about the structure and dynamics of the magnetosphere (e.g., the hardness variations suggest some temperature stratification of the hot plasma\cite{2016AdSpR..58..680U}). This information is needed to test models that can be used to constrain mass-loss rates of populations of magnetic massive stars. 

The X-ray properties of magnetic OB stars differ in several significant ways from those of non-magnetic, single O stars. Theory predicts that the levels of X-ray emission from Magnetically Channeled Wind Shocks (MCWS) in OB stars should exceed the levels from the embedded wind shocks seen in non-magnetic massive stars and that the X-ray emission should be harder\citep{2016AdSpR..58..680U}. This follows from the additional shock-heating resulting from the collision and eventual merging of initially separately channeled wind streams. Additionally, X-ray emission lines should be significantly narrower, due to the magnetic confinement. {\it Chandra} and {\it XMM-Newton} grating observations of only a few magnetic OB stars show X-ray lines that tend to be narrow, though frequently somewhat broader than simple models predict \citep{2005ApJ...628..986G}. In addition, a diversity of behaviors has been detected: only some magnetic OB stars have especially hard emission, perhaps due to shock retreat or to mixing of hot and cool gas on timescales shorter than the cooling time; and the location of the hot plasma is apparently incompatible with the observed variations if they are due to occultation as expected; brighter than expected X-ray emissions were seen in some cases\cite{2016AdSpR..58..680U,2014ApJS..215...10N}.

{\it LEM} will provide unprecedented spectral detail for high signal-to-noise observations while still providing {\em Chandra}- and {\it XMM-Newton}-grating-quality spectra for numerous more distant and fainter objects. Critical diagnostics will include line widths and profiles to diagnose plasma dynamics and the degree of magnetic confinement, temperature diagnostics from SED modeling, and helium-like line ratio diagnostics that, in hot stars, are sensitive to the location of the X-ray emitting plasma with respect to the photosphere (Sect.~\ref{s:helikes})\cite{Kahn2001,Leutenegger2006,2016ApJ...831..138N}. These X-ray spectral diagnostics will constrain models of magnetospheric physics and mass-loss from these stars. 

In addition to the confined winds, it has recently been suggested that X-rays in some late B or A stars are due to auroral emissions linked to material falling back onto the star after magnetic reconnection at the edge of the magnetosphere. Only a few examples were studied up to now\cite{2018A&A...619A..33R,2017MNRAS.467.2820L,2020MNRAS.493.4657L}, and none at high resolution. {\it LEM} can provide a breakthrough as studying the lines at high-resolution will allow us to pinpoint their exact origin.
\vspace{0.5cm}

\section{WOLF-RAYET STARS}
\label{s:wr}

\begin{figure*}[t] 
   \centering
   \includegraphics[width=0.4\textwidth]{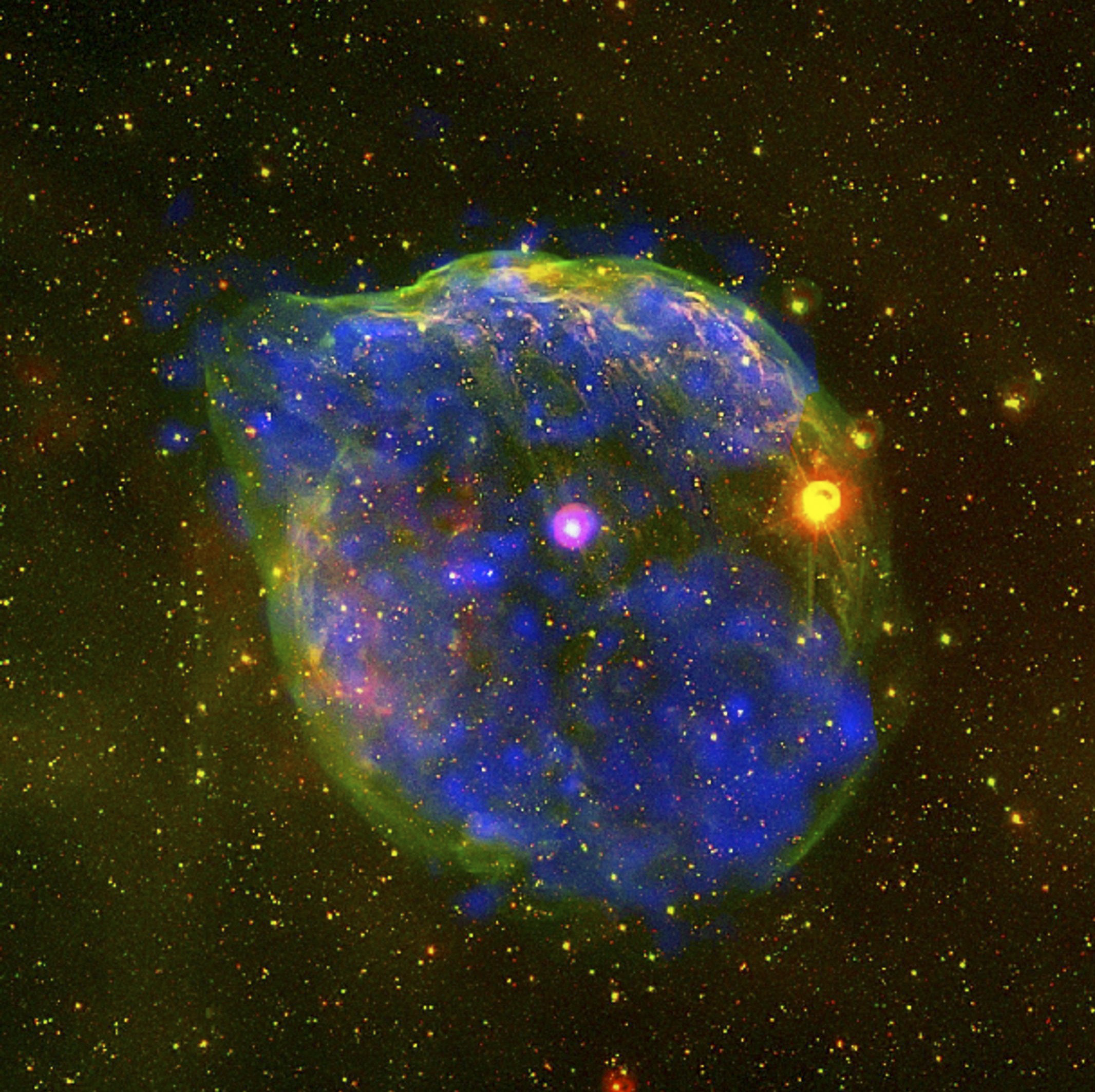} 
    \includegraphics[width=0.55\textwidth]
    {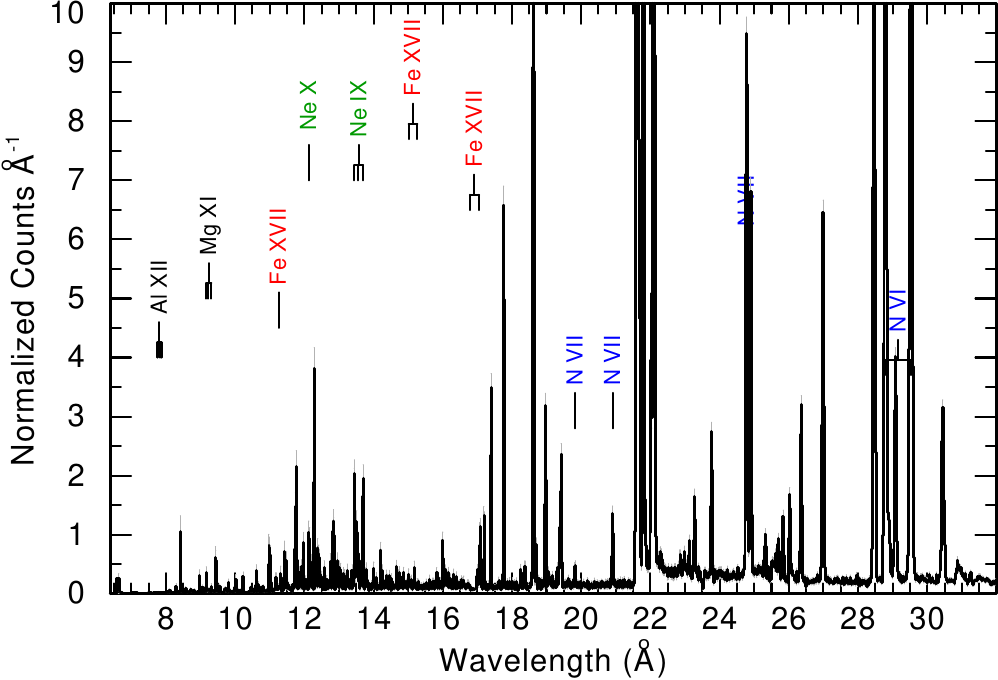}
         \caption{\textbf{Left}: 
         Based on Figure\,4 from Ref~\citenum{Toala2012} showing a composite color picture of the {\it XMM-Newton} EPIC image (blue) and CTIO [O\,{\sc iii}] (green) and H$\alpha$ (red) images of the S\,308 nebula which is blown by the central WR star, WR6.  Both the nebula and the star are X-ray sources. The image size is  $20'\times 20'$, fitting nicely within the {\it LEM} $30'\times 30'$ field of view. \textbf{Right}: {\it LEM} spectrum of the S\,308 nebula; the simulation is done assuming the measured X-ray flux and assuming a 10\,ks exposure.}
   \label{fig:wr6}
\end{figure*}

Wolf-Rayet (WR) stars are among the key drivers of stellar feedback in galaxies. These hot and compact stars are hydrogen-depleted descendants of the most massive stars on the evolutionary stage shortly before the core collapse event \citep{Crowther2007}. Likely, WR stars are the primary suppliers of stellar mass black holes. A significant fraction of WR stars are members of binary systems and are likely among the progenitors of gravitational wave sources.  Being hot and compact, WR stars drive the strongest stellar winds ($\dot{M} \gtrsim 10^{-5}$~M$_{\odot}$~yr$^{-1}$) among all types of non-degenerate stars. The lifetimes of WR stars are relatively short, $\sim 10^5$\,yr, but the energy supplied to the ISM by the WR winds during this time is comparable to that of a SNe.
Many WR stars blow spectacular bubbles filled with hot plasma (Fig.\,\ref{fig:wr6}). 

Early {\em Einstein} X-ray telescope observations revealed that binary WR stars are usually significantly brighter in X-rays compared to single WR stars \citep{Pollock1987}. In such systems,
the bulk of the X-ray emission is attributed to the collision of winds from O and WR binary components. The  WR colliding wind binaries are typically (although not always \cite{2021MNRAS.501.4214N}) bright X-ray sources that usually show  variability  explained by orbital
motion (see section \ref{sec:cwb}). In some objects,  besides X-rays,  non-thermal radio emission
and dust are  produced \citep{Pittard2006}.

Single WR stars are poorly studied in X-rays. There are as yet no consistent models capable of explaining the production of X-rays in these objects \citep{Gayley2016}. The problem is that 
the winds of WR stars are so dense that X-rays would be expected to be completely absorbed within the wind and not be observed, unless they can be formed further out where the wind density is lower. X-rays in WR winds thus require mechanisms that can produce hot gas at much larger radii than in lower density OB winds. 

Despite the large astrophysical significance of WR stars,  up to now, high-resolution X-ray spectra of only one single WR star have been obtained \citep{Oskinova2012, Huenemoerder2015}, and no high-resolution X-ray spectra of WR bubbles exist. {\it LEM} will obtain the first ever high-resolution X-ray spectra of WR bubbles, and deliver the first sample of X-ray spectra of WR stars which is needed to understand how X-rays are generated and how X-rays affect their stellar winds. This will revolutionize our understanding
of the energetics of WR stars and their winds.

Analyses of low-resolution X-ray spectra of WR stars indicate the presence of plasma  with  temperatures  between 1\,MK up to 50\,MK distributed throughout the stellar wind. It appears that X-ray emission is probably due to the line-driving instability as in O type stars. 
A handful of WR stars that were monitored in X-rays show significant variability thought to be on the rotational time scale \citep{2011A&A...527A..66G,Ignace2013,2011ApJ...735...13H}. The reasons for this variability are not yet known, but might be due to coherent, large scale density structures in the wind. {\it LEM} will obtain high-quality spectra in sufficiently short exposures to allow us to correlate spectral and light-curve variability to help reveal the nature of the underlying variability.  

\begin{figure*}[t] 
   \centering
   \includegraphics[width=0.49\textwidth]{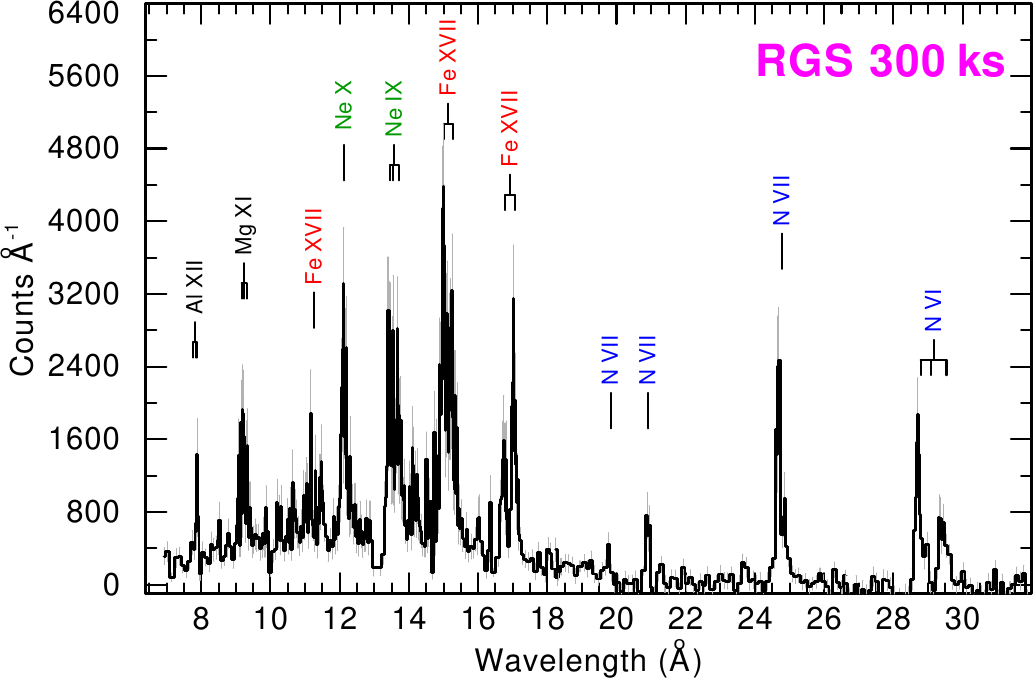} 
    \includegraphics[width=0.49\textwidth]
    {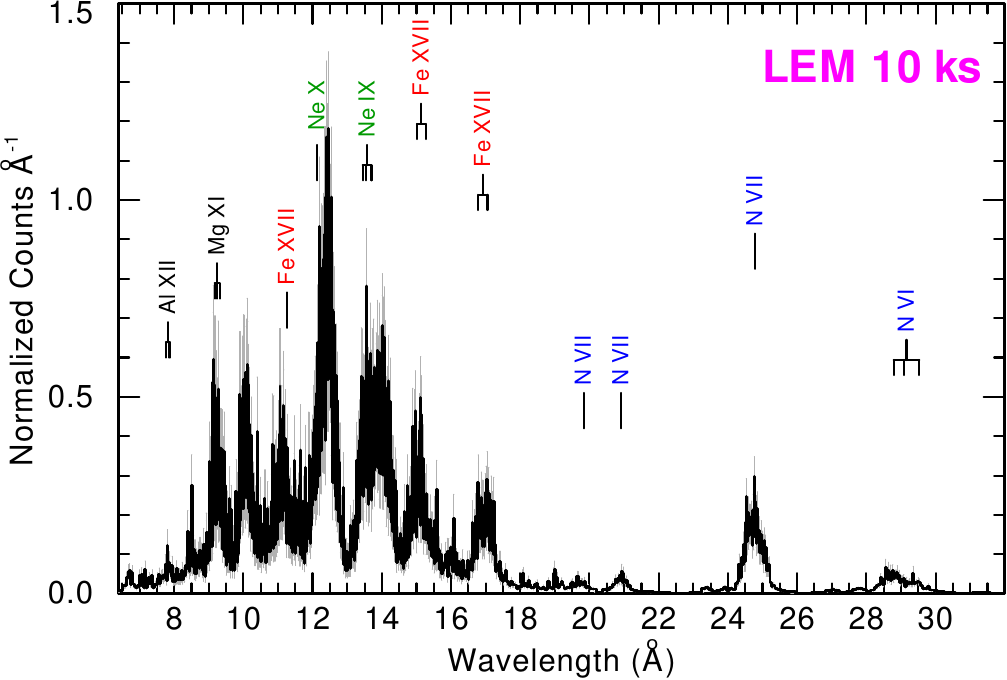}
         \caption{\textbf{Left}: 
         Observed {\it XMM-Newton} RGS spectrum of WR\,6, the central star of the S\,308 nebula (Fig.\,\ref{fig:wr6}) \citep{Oskinova2012}. The spectrum was obtained with a 
         400\,ks exposure. The lower signal-to-noise {\it Chandra} spectrum of the same source was collected with a 600\,ks exposure time \citep{Huenemoerder2015}.
         \textbf{Right}: Simulated {\it LEM} spectrum of the WR\,6 nebula compouted assuming the measured X-ray flux and a 10\,ks exposure.}
   \label{fig:rgslem}
\end{figure*}

WR stars and their winds are, in general, hydrogen free but strongly enriched in CNO and other metals. Soft X-ray spectroscopy is especially well suited for studies of these metal rich winds and the bright H-like and He-like line complexes of C, N and O should be easily detected in {\it LEM} spectra. X-ray spectra also provide excellent abundance diagnostics of elements such as sodium \citep{Huenemoerder2015} which can be  difficult to  obtain  at other wavelengths.

The high spectroscopic resolving power of {\it LEM} will be used to resolve the emission lines, including those of He-like ions that can diagnose the strength of the local FUV radiation field (Sect~\ref{s:helikes}), and determine the detailed shape of X-ray lines in a sample of WR stars, of both low and high masses and different WR subtypes. Using  existing modeling techniques \citep{Leutenegger2006, Shenar2015},  these will provide important constraints on the locations of the  X-ray emitting plasma. It is also possible that absorption edges in the X-ray spectra of WR stars can serve as independent mass-loss diagnostics.

The chemical abundances in WR winds will be measured to see how abundances in the hot plasma depend on the evolutionary status of the WR star, and will also be compared with 
nebular abundances of the WR bubbles to provide a history of chemical enrichment.  High resolution X-ray spectra of the bubbles themselves will reveal the non-thermal velocity dispersion and determine what are the abundances and energetics of hot bubbles around massive stars.

\begin{figure*}[ht] 
   \centering
   \includegraphics[width=1.0\textwidth]{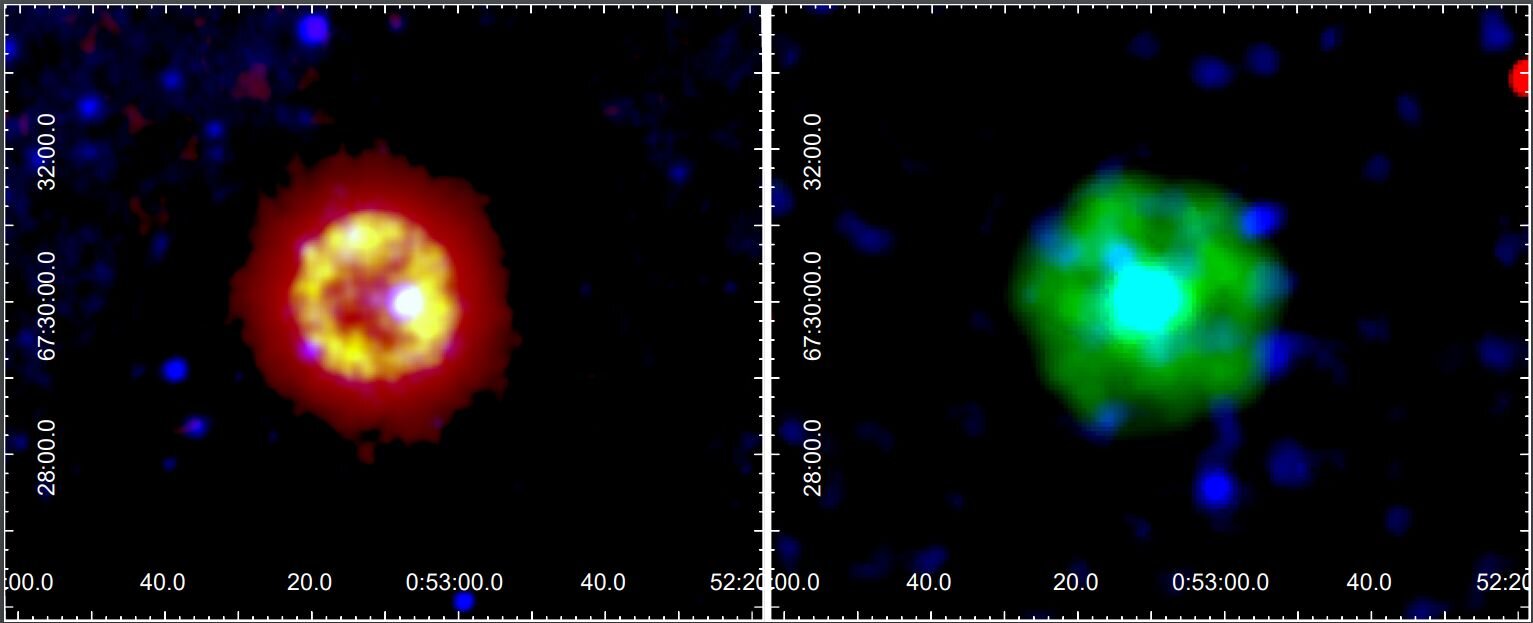} 
         \caption{
         Images of IRAS 00500+6713 in mid-infrared (IR) and X-ray wavelength ranges. \textbf{Left:} IR Wide-field Infrared Survey Explorer (WISE) image. \textbf{Right:} X-ray XMM-Newton EPIC image: red: 0.2-0.7 keV, green: 0.7-1.2 keV, blue: 1.2-7.0 keV. The adaptively smoothed image shows that X-ray emission uniformly fills the whole extent of the IR nebula. The coordinates are in units of RA (J2000) and Dec (J2000) on the horizontal and vertical scales, respectively. Adopted from \citep{Oskinova2020}.}
   \label{fig:iras}
\end{figure*}

Among WR stars, there is a group of objects originating from lower mass stars, such as the central stars of planetary nebulae which show WR-type spectra in optical wavelengths, or white dwarf merger products which have super-Chandrasekhar masses. An example of the later is IRAS\,00500 (Fig.\ref{fig:iras}). Objects such as IRAS\,00500 have extremely powerful stellar winds, are hydrogen and helium free, and are strong X-ray sources \citep{Gvaramadze2019, Oskinova2020}. Yet, this important class of objects has never been studied in X-rays with high-resolution. A {\it LEM}-class mission with order of magnitude greater sensitivity than {\it Chandra} or {\it XMM-Newton}
is required for this task.  

\section{COLLIDING WIND BINARIES}
\label{sec:cwb}

In addition to the X-ray emitting shocked gas intrinsic to and embedded within radiatively-driven stellar winds, massive stars in binary systems also produce X-ray emission by the collisions of their stellar winds in the space between the stars \citep{1976SvAL....2..138C, 2008MNRAS.388.1047P}.  The "colliding wind" emission is localized along a conical shock surface which is determined by the relative wind momentum fluxes (mass loss rate $\times$  wind velocity) of the binary components.  The thermal X-ray emission produced in the colliding wind shock provides unique information on the wind density, wind speed, and abundance of the wind along the surface of the shock, and can also be used to test the assumption of spherical symmetry, without biases of clumping (if the cooling is mainly adiabatic).

A particularly interesting case of an X-ray emitting colliding wind binary is the star \etac\ \citep{1997ARA&A..35....1D}, a binary system that contains the only star near 100\ms\ \citep{2001ApJ...553..837H} within 3~kpc.  The primary star is thought to be in the short-lived, unstable Luminous Blue Variable phase.  The star is well-known for its "Great Eruption" in the 19th century \citep{hersch1868}, an event nearly as energetic as a supernova, and is the prototype of the class of "supernova imposters".  It is currently unclear if this event was driven by instabilities intrinsic to the star, or by an extrinsic merger between two stars in a triple system.  

\begin{figure}[htbp] 
   \centering
 \includegraphics[width=1.0\columnwidth]{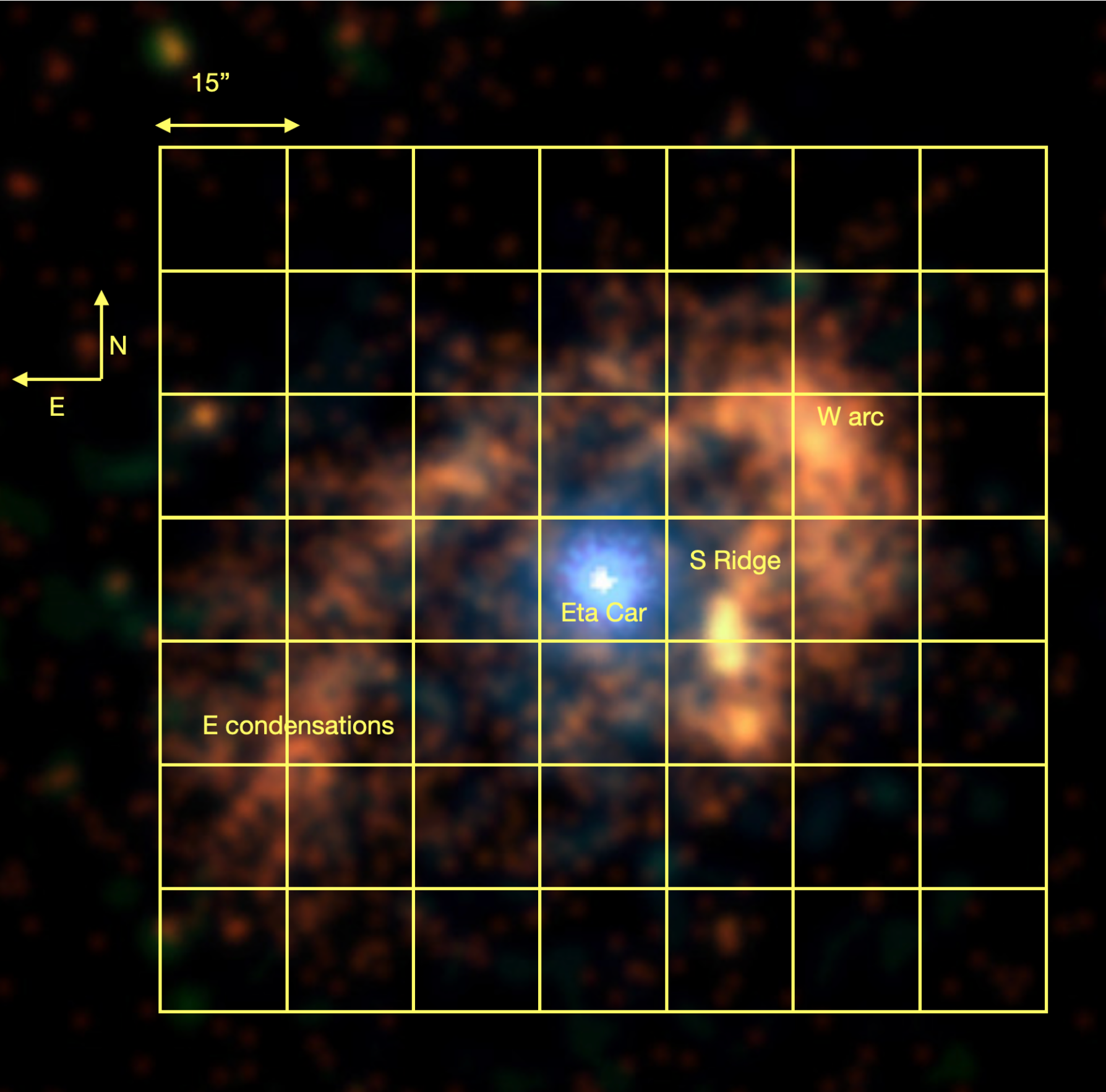} 
   \caption{{\it Chandra} X-ray image of the \etac\ colliding wind system (in blue white at the center of the image) and the X-ray bright stellar ejecta (in gold) which surrounds it.  The central $7\times7$ {\it LEM} pixel array, centered on \etac, is superimposed.  {\it LEM} will provide 1~eV resolution spectra in the 0.2--2.0 keV band of the colliding wind binary and the entire ejecta field surrounding it (see Figure\ref{fig:etasims}).}
   \label{fig:cheta}
\end{figure}

The colliding wind binary system is surrounded by thick ejecta produced by the Great Eruption (and earlier, more minor ejections) which reflect and produce shocked thermal X-ray emission \citep{1979ApJ...234L..55S, 2004ApJ...613..381C, 2022ApJ...937..122C}.  High-resolution, spatially resolved X-ray spectra from {\it LEM} of \etac\ (and other colliding wind binaries) will resolve long outstanding questions about the wind density profile along the colliding wind shock through constraining line density diagnostics of the shocked gas and measurement of absorption edges produced by the  intervening unshocked stellar wind as they vary around the orbit (see models of \cite{2021A&A...646A..89M}). Figure \ref{fig:cheta} shows an example {\it LEM} observation of \etac, where \etac\ itself is  in the central {\it LEM} pixel with the X-ray ejecta field covered by the central  $7\times7$ pixel array providing a spectral resolution of 1~eV. Also shown is a simulated {\it LEM} spectrum of the heavily-absorbed colliding wind emission and (mostly unabsorbed) emission from the shocked ejecta. 

\begin{figure}[htbp] 
   \centering
\includegraphics[trim={0 1.0 0 1.0cm},clip,width=1.05\columnwidth]{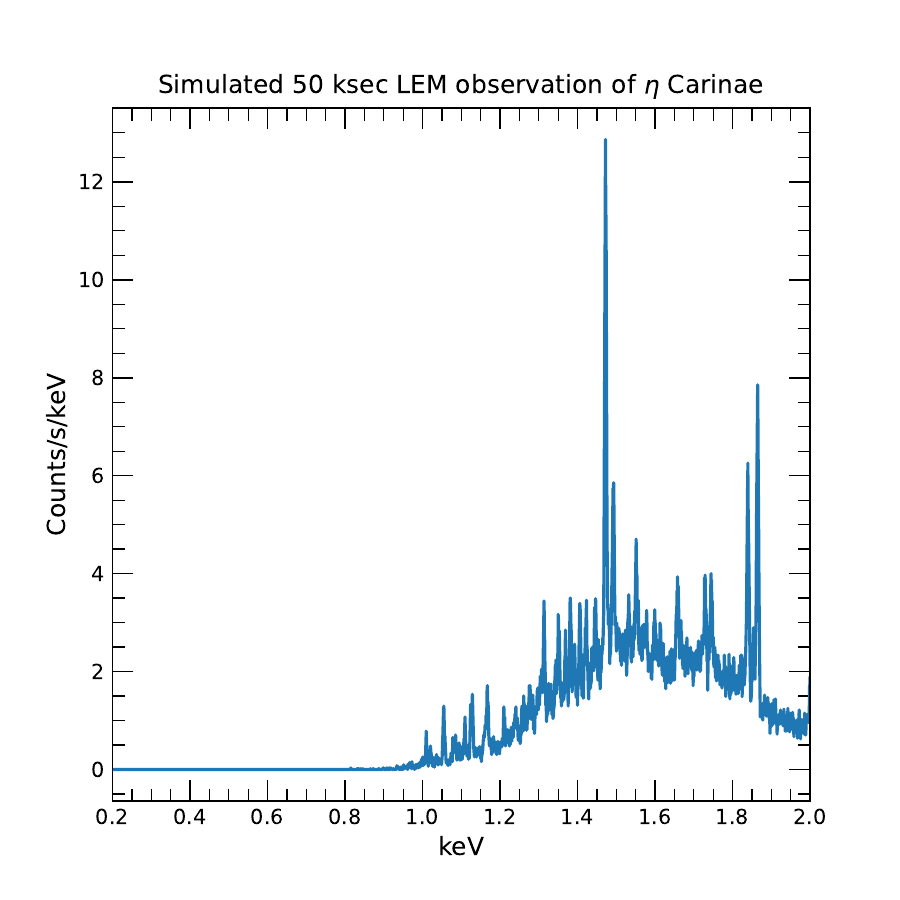}
\includegraphics[trim={0 1.0 0 1.0cm},clip,width=1.05\columnwidth]{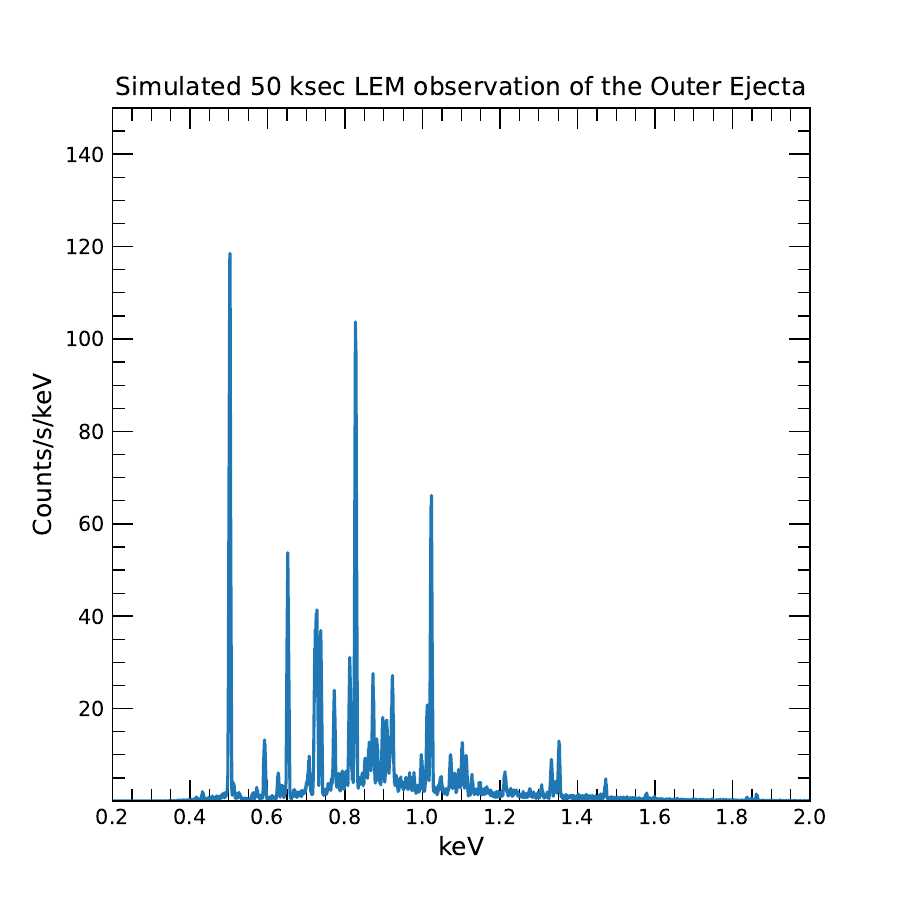}
\caption{
   \textbf{Top}: Simulated 50~ksec {\it LEM} spectrum of the \etac\ colliding wind emission. This emission has a temperature of 1--4~keV typically and is heavily absorbed ($N_{H}> 3\times10^{22}$ cm$^{-2}$ which suppresses emission below 1 keV.  \textbf{Bottom}:  Simulated  50~ksec {\it LEM} spectrum of the X-ray emission from the outer ejecta.  This material is unabsorbed with a typical temperature of 0.5~keV and very N-enhanced.}
   \label{fig:etasims}
\end{figure}

{\it LEM} will break new ground in our understanding of the historical record of the outbursts from \etac\ (and other eruptive Luminous Blue Variables, LBVs) by measuring temperature, velocity, and abundance variations in the outer ejecta, ejected from 1200 CE to 1843 CE \citep{Kiminki:2016kq}.  {\it LEM} spectra will also explore the importance of charge-exchange emission between the ionized and neutral gas near the collisionless shocks.  {\it LEM} spectra will provide the definitive  physical parameters of the shocked winds and ejecta better than any previous instrument, and improve our understanding of (quasi-)stable mass loss from stars in evolved binaries by more than an order of magnitude.

Beyond \etac, {\it LEM} will be able to study at high-resolution a large sample of colliding wind binaries. Only a few systems have been observed by current facilities\cite{2008A&A...490..259S,2016A&A...589A.121R,2022MNRAS.513.6074M,2021ApJ...915..114P}. X-ray line profiles recorded for massive colliding wind binaries is a powerful probe into the stellar winds. Indeed, the profiles only depend on the system geometry (which is known) and stellar winds properties. Self-consistent computation of line profiles\cite{2016NewA...43...70R,2021A&A...646A..89M} are available and show the sensitivity of the profiles hence their diagnostic value in this context. [the database can be used for simulations!]

\section{Be STARS}
\label{s:bestars}

Be stars, or B-type emission-line stars, are a class of rapidly-rotating B-type main-sequence or slightly evolved stars that exhibit prominent or periodic Balmer emission lines in their spectra.\cite{Porter2003} The emission lines arise from reprocessing of photospheric radiation by the presence of a surrounding disk of material ejected from the oblate star by a combination of radiation pressure and centrifugally-reduced effective gravity.\citep{Struve1931,Massa1975} The strengths and shapes of the emission lines can vary over time, reflecting variations in mass transfer rates from star to disk, resulting disk properties, and star-disk interaction.\cite{Porter2003}

The X-ray properties of Be stars are generally similar to those of single OB stars, except for perhaps being somewhat brighter.\cite{Cohen2000} Consequently, the X-ray generation mechanism is attributed to similar line-driven wind shocks. However, the presence of the circumstellar disk could also give rise to hot shocked gas as a result of wind-disk interaction, and there is also evidence for magnetic fields playing a role in some systems.  High-resolution {\it LEM} spectra will be able to provide density and ambient UV radiation field diagnostics through the He-like ions. 

It should also be noted that Be stars are more numerous at low metallicities hence the knowledge of their feedback becomes increasingly important.

Be stars in high-mass X-ray binaries, in which X-rays are produced predominantly through accretion of disk and wind material onto a neutron star companion, also present interesting insights into the Be star phenomenon,\cite{Coe2000} although this topic is beyond the scope of the present paper.

Amongst the category of single Be stars, some display a bright and hard X-ray emission of debated origin. Such objects are called "$\gamma$ Cas analogs", from their prototype. 
These objects are characterized by thermal, 5--15\,keV emission, typically showing  prominent emission from the iron complex near 7\,keV \cite{Smith2016}, along with emission from plasma at lower temperatures\cite{2004ApJ...600..972S}. The origin of these emissions is unclear, although they are certainly linked to the disks: either a companion is involved in the X-ray generation\cite{Postnov2017, Naze2022, Gies2023}, or star-disk interactions are responsible\cite{1999ApJ...517..866S}. Preliminary  results from line profile modelling (Rauw et al., in prep) indicate large changes of the line profiles depending on the chosen emission model. 

The $\gamma$ Cas analogues are
variable X-ray sources\cite{Smith2016,2016ApJ...832..140H,2022A&A...664A.184R},  and the reason for this X-ray variability is not fully known either.  High-resolution X-ray spectra of the $\gamma$-Cas analogues from {\it LEM} will help constrain 
the origin of the hot gas and the reasons for its variability.

{\it LEM} will thus be instrumental in solving a mystery that affects about 10\% of Be stars\cite{2023MNRAS.525.4186N,2018A&A...619A.148N}.


\section{X-RAY EMISSION FROM HOT STARS IN OTHER GALAXIES}
\label{s:othergals}

Massive stars are relatively faint and soft X-ray sources. As a rule of thumb, they emit only $\sim 10^{-7}$ of their bolometric luminosity in X-rays, while the temperature of X-ray emitting plasma is only $\sim~5$\,MK. Even the X-ray brightest massive stars, such as colliding wind binaries,  have luminosities smaller than $< 10^{34}$\,erg\,s$^{-1}$ in the 0.2-12.0\,keV band \citep{Nebot2018}. This corresponds to fluxes $< 10^{-14}$\,erg\,s$^{-1}$\,cm$^{-2}$ for stars in the Magellanic Clouds galaxies, and 
$< 10^{-16}$\,erg\,s$^{-1}$\,cm$^{-2}$ in the M\,31 galaxy. 
Corresponding count rates for the high-resolution spectrograph on board the largest modern X-ray telescope, {\it XMM-Newton}, are $<10^{-3}$\,s$^{-1}$. Therefore, it is not surprising that no high-resolution X-ray spectrum of a massive star outside of our Galaxy exists.  

{\it Chandra} observations of the Tarantula nebula in the Large Magellanic Cloud (LMC) galaxy ($d\approx 50$\,kpc) is the deepest X-ray probe of massive stars in nearby galaxies to date \citep{Crowther2022}. The Tarantula nebula region, encompassing the 30 Dor star forming complex, contains hundreds of O and WR-type stars, hence allowing meaningful comparison with the samples of Galactic massive stars. These Chandra observations revealed that X-ray properties of O and WR stars in the LMC are very similar to their counterparts in the Galaxy, despite the fact that stellar winds of metal-poor stars in the LMC are somewhat weaker compared to Galactic massive stars. This confirms previous hints from the observations of the giant H{\sc ii} regions N11 and N206 \cite{2012A&A...547A..19K,2018ApJ...853..164N, 2018A&A...615A..40R}.

 \begin{figure}[ht] 
   \centering
   \includegraphics[width=1.0\columnwidth, 
   angle=-90]{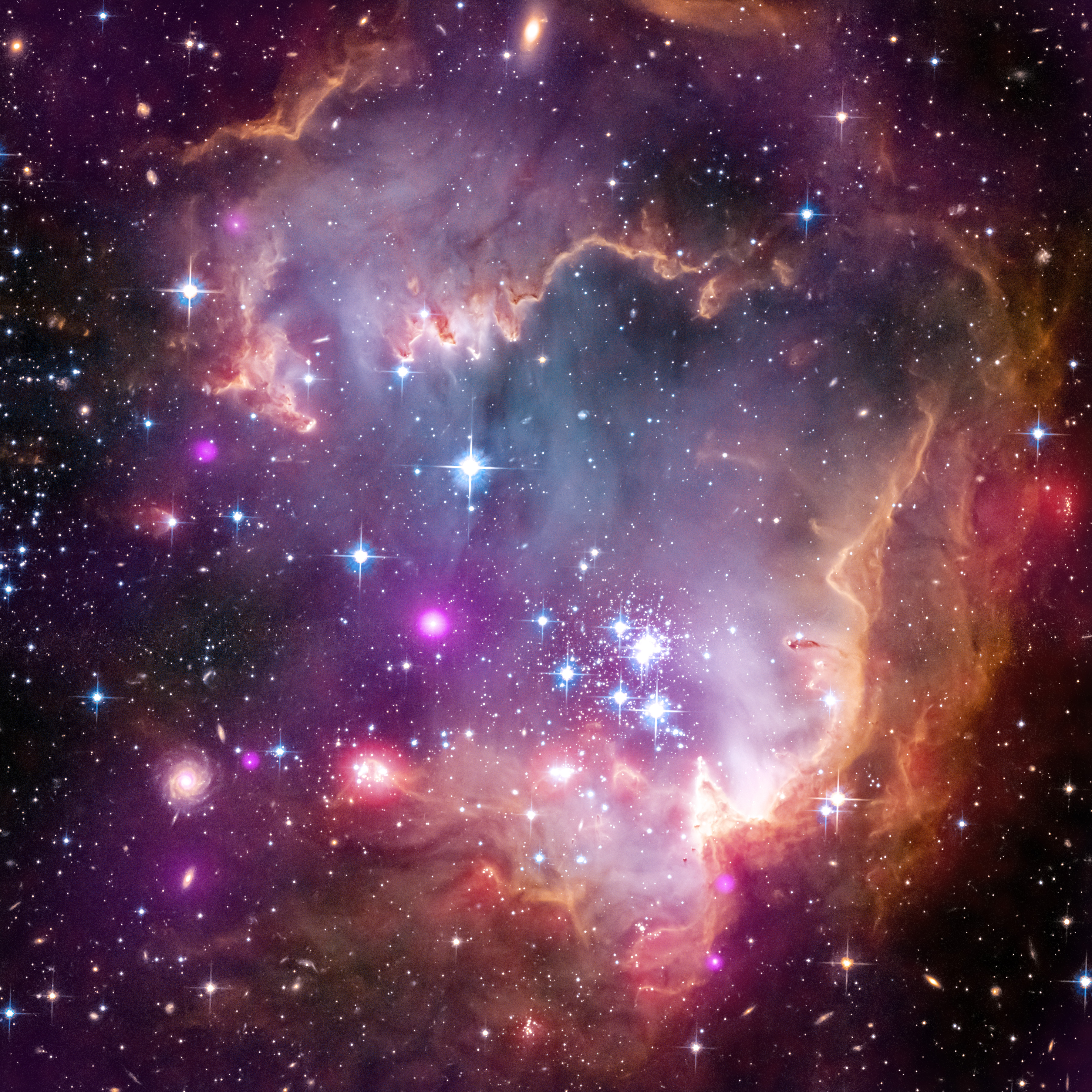} 
         \caption{Composite image of the NGC\,602 young massive star cluster in the SMC. The \chandra\ data are shown in purple; visible light seen by the NASA/ESA {\it Hubble Space Telescope} is in red, green, and blue; and infrared data from the {\it Spitzer Space Telescope} are coloured red.  The brightest X-ray point sources are background AGN. The diffuse X-ray emission in which O-type stars are embedded is due to the unresolved population of X-ray active low-mass protostars and pre-main sequence stars.  The image size is $\approx 2'\times 2' (\approx 32 \times 32\,{\rm pc})$.
X-ray: NASA/CXC/Univ.Potsdam/L.Oskinova et al; Optical: ESA, NASA/STScI; Infrared: NASA/JPL-Caltech
         }
   \label{fig:ngc602}
\end{figure}

The Small Magellanic Cloud (SMC) ($d\approx 62$\,kpc) has even lower metallicity than  the LMC.  The X-ray brightest massive star in the SMC is the remarkable colliding wind binary HD\,5980, which contains an eruptive LBV\citep{Naze2007} and an O star (and a third, more distant, less massive companion).  The object shows periodic modulations of its X-ray flux likely related to the orbital motion. These modulations have been seen to change over the last two decades, demonstrating the existence of a theoretically predicted phenomenon linked to thin-shell instabilities \citep{Naze2018}. This renders the system unique. High-resolution X-ray spectroscopy with {\it LEM} will provide important physical constraints on the nature of complex wind interactions in this system (and others like it).  These massive binaries are the probable progenitors of binary black hole gravitational wave sources. 

There is some tentative evidence that O-type stars in the SMC are less X-ray luminous than the higher metallicity O stars in the Galaxy and the LMC \citep{Oskinova2013}. Even Sk~183 (O3V), the most massive star in the NGC\,602 cluster, was not detected by deep {\it Chandra} observations, placing an upper limit on its X-ray luminosity of $L_{x}<10^{32}$\,erg\,$^{-1}$. On the other hand, {\it Chandra} observations of such compact clusters revealed an unresolved population of pre-main-sequence massive stars and established that diffuse X-ray emission in young clusters, such as NGC\,602, is {\em not} always 
produced by a hot bubble blown by stellar winds (Fig.\,\ref{fig:ngc602}).  

{\it LEM} will be unprecedented in its ability to study young star clusters in other galaxies. Thanks to the {\it Chandra} high spatial resolution observations of nearby galaxies, we are now able to correctly model contributions from various constituencies of young, massive clusters---low-mass stars, the hot phase of the interstellar medium, and individual massive stars---which is necessary for unveiling the full potential of {\it LEM} spectroscopy.  An especially promising avenue are joint {\it LEM} and JWST explorations of star forming regions in galaxies. 

Complemented by the JWST observations in the infra-red, and learning from the {\it Chandra} and {\it XMM-Newton} legacy, {\it LEM} will for the first time allow detailed studies of metal-poor massive stars and their feedback in the nearby dwarf galaxies. In particular, {\it LEM} spectra will determined what the properties of metal-poor massive stars are whether WR stars, LBVs, and colliding wind binaries in other galaxies have similar X-ray properties to their counterparts in the Galaxy. If not, {\it LEM} will help answer how these differences can be explained and what it tells us about stellar wind driving at different metallicities.

\section{EXOPLANETS AROUND HOT STARS}
\label{s:exoplanets}

At the time of writing, the NASA Exoplanet Archive contained 5528 confirmed planets.\footnote{https://exoplanetarchive.ipac.caltech.edu/} While the search for planets will continue unabated, the scientific emphasis has pivoted in recent years from planet detection to understanding planet demographics, formation, evolution, and the prospects for habitability and harboring life. The importance of host star EUV and X-ray radiation has come to prominence on realisation that energetic coronal emission can deplete and evaporate the envelopes and atmospheres of planets, strongly influencing the potential for habitability and in some cases possibly resulting in their entire erosion\citep[e.g.][]{Sanz-Forcada+2011, OwenWu2013}.

The occurrence rate of planets is now known to increase with stellar mass, peaking close to $1.9\,M_\odot$ (SpT$\sim$A5), but then dropping precipitously for spectral types earlier than A \citep{Reffert+2015, Johnson+2010}. This might not be unexpected since the intense UV and X-ray radiation fields are efficient at dispelling cirumstellar gas and in dissipating protoplanetary disks\citep{GortiHollenbach2009}. Indeed,
the disk phase in more massive stars is much more brief than for low mass stars, and it is not clear if planets have sufficient time to form before the disk and circumstellar material is gone.

However, recent detections of planets around hot stars such as b Centauri (AB)b \citep{Janson+2021}, Mu2 Scorpii b \citep{Squicciarini+2022} or V921 Sco b \citep{UbeiraGabellini+2019}, indicate that the formation of planets, even gas-rich ones, around hot stars is indeed possible. As in the case of low-mass stars, the EUV and soft X-ray radiation fields of higher mass stars are expected to be key to understanding the erosion or survival of gas-rich planets. The soft X-ray response of {\it LEM} can help in this regard, probing spectral regions where lower temperature plasma and plasma diagnostics contribute and leading to a better understanding of the EUV-X-ray emission.

\section{CONCLUDING REMARKS}

We have laid out some of the salient aspects of the scientific requirement for high resolution soft X-ray spectroscopy of early-type stars. While the {\it Chandra} and {\it XMM-Newton} diffraction gratings have already revolutionized our view of hot star physics, the limitations of their capabilities have been reached and more sensitive next generation instruments are needed to make the next breakthroughs. 

The major strength of {\it LEM} for hot star physics is its large effective area, aided by the inherent energy resolution of its microcalorimeter that readily achieves resolving powers of 1000 and obviates the need for relatively inefficient dispersive optical elements.  This increased sensitivity enables much fainter and more distant objects to be observed, greatly increasing the pool of potential targets. For brighter sources, the sensitivity opens up time domain studies, wherein sufficient signal can be garnered in short order and exposure times, probing source variations on ks timescales. 

We have argued that these capabilities of {\it LEM} will yield breakthroughs in all types of hot star systems, from understanding single OB and WR star winds and how they vary with metallicity, to probing the shocks of colliding wind systems and the magnetically channeled winds of magnetic OB stars. {\it LEM} will also study the energetics of WR star bubbles and feedback from their powerful pre-SN stellar winds.

\bibliography{lit}
\bibliographystyle{aasjournal}

\end{document}